\documentclass{nature-pre}
\usepackage{times}

\usepackage{graphicx}
\usepackage{amsmath}
\usepackage{amssymb}
\usepackage{multirow}

\graphicspath{./}

\title{Symmetric carbon tetramers forming chemically stable spin qubits in hBN}
\author{Zsolt Benedek$^{1,2,3}$, Rohit Babar$^{1,2}$, \'{A}d\'{a}m Ganyecz$^{1,2}$, Tibor Szilv\'{a}si$^{3}$, \"Ors Legeza$^1$, Gergely Barcza$^{1,2,3,*}$, Viktor Iv\'{a}dy$^{2,4,5,*}$} 

\begin{document}
\maketitle

\begin{affiliations}
\item  {Strongly Correlated Systems Lend\"{u}let Research Group, Wigner Research Centre for Physics, PO Box 49, H-1525, Budapest, Hungary}
\item {MTA-ELTE Lend\"{u}let "Momentum" NewQubit Research Group, P\'{a}zm\'{a}ny P\'eter, S\'et\'{a}ny 1/A, 1117 Budapest, Hungary}
\item{Department of Chemical and Biological Engineering, The University of Alabama, Tuscaloosa, Alabama 35487, United States}
\item{Department of Physics of Complex Systems, E\"otv\"os Loránd University, Egyetem t\'er 1-3, H-1053 Budapest, Hungary}
\item   {Department of Physics, Chemistry and Biology, Link\"oping University, SE-581 83 Link\"oping, Sweden}
\item[*] email: barcza.gergely@wigner.hu, ivady.viktor@ttk.elte.hu
\end{affiliations}

\date{\today}

\newpage

\begin{abstract}
Point defect quantum bits in semiconductors have the potential to revolutionize sensing at atomic scales.  Currently, vacancy related defects, such as the NV center in diamond and the VB$^-$ in hexagonal boron nitride (hBN), are at the forefront of high spatial resolution and low dimensional sensing. On the other hand, vacancies' reactive nature and instability at the surface limit further developments. Here, we study the symmetric carbon tetramers in hBN and propose them as a chemically stable spin qubit for sensing in low dimensions. We utilize periodic-DFT and quantum chemistry approaches to reliably and accurately predict the electronic, optical, and spin properties of the studied defect. We show that the nitrogen centered symmetric carbon tetramer gives rise to spin state dependent optical signals with strain sensitive intersystem crossing rates. Furthermore, the weak hyperfine coupling of the defect to their spin environments results in a reduced electron spin resonance linewidth that may enhance sensitivity.
\end{abstract}

\newpage


\section*{Introduction}
Condensed matter physics in low dimensions is already a vast, yet rapidly growing field. Especially, transition metal dichalcogenides\cite{wang_electronics_2012,manzeli_2d_2017} and complex van der Waals heterostructures\cite{geim_van_2013,liang_van_2020} develop with unprecedented pace. The study of these nanometer-scale structures and related phenomena demands novel high-spatial resolution sensing devices  operating in a wide temperature range and sensitive to various external fields. Point defect quantum bits in semiconductors, such as the NV center in diamond\cite{Jelezko:PRL200492,DohertyNVreview} and the silicon vacancy in silicon carbide\cite{Widmann2014}, have already demonstrated outstanding capabilities in high-spatial resolution sensing and fulfilled many of these requirements.\cite{taylor_high-sensitivity_2008,schirhagl_nitrogen-vacancy_2014,barry_sensitivity_2020,sturner_integrated_2020,zhang_toward_2021} The distance of the sensor from the targeted system is of crucial importance for high-spatial resolution sensing. Therefore, further improvements require point defect sensors to be engineered closer to the surface or to be directly integrated into various low-dimensional structures. The currently available point defect qubit sensors in 3D semiconductors are not optimal for such applications due to their inherently bulk nature and strong dependence on surface chemistry.\cite{kaviani_proper_2014,kim_decoherence_2015,dwyer_probing_2022} The development of point defect qubits in layered van der Waals semiconductors may provide a way to overcome this obstacle as their surface is chemically stable and the thickness of the host material, and thus the distance of the qubits and the surface, can be engineered straightforwardly by exfoliation.\cite{tetienne_quantum_2021,healey_quantum_2022,kumar_magnetic_2022} Furthermore, van der Waals semiconductors with spin qubits can implement atomic thin sensors with advanced capabilities.\cite{gottscholl_spin_2021,liu_temperature-dependent_2021,tetienne_quantum_2021,healey_quantum_2022,kumar_magnetic_2022,lyu_strain_2022}

Hexagonal boron nitride (hBN) is a layered wide-band gap semiconductors, which is often used in van der Waals heterostructures. Its large, close to 6~eV band gap accommodates numerous optically active electronic states of structural defects and impurities.\cite{caldwell_photonics_2019,sajid_single-photon_2020} Exfoliated hBN samples may contain point defects in such a low number that even individual color centers can be observed with confocal microscopy techniques. Numerous single photon emitters were demonstrated in hBN that has opened a new field.\cite{tran_quantum_2016,caldwell_photonics_2019} Point defect quantum bits form a special class of color centers that carry high spin ground and optically excited states and feature a spin dependent optical emission. This phenomena makes optical detection of magnetic resonance (ODMR) measurements possible. ODMR signal of different spin qubits have already been reported\cite{gottscholl_initialization_2020,chejanovsky_single-spin_2021,mendelson_identifying_2021,stern_room-temperature_2022} and predicted\cite{sajid_vncb_2020,babar_quantum_2021,bhang_first-principles_2021,liu_spin-active_2022} in hBN.  One of the observed ODMR centers have been identified as the negatively charged boron vacancy center (VB$^-$ center).\cite{gottscholl_initialization_2020,ivady_ab_2020,haykal_decoherence_2022,liu_coherent_2022}  Identification of other ODMR centers remains elusive, despite the numerous experimental\cite{mendelson_identifying_2021,chejanovsky_single-spin_2021,stern_room-temperature_2022} and theoretical works\cite{jara_first-principles_2021,li_carbon_2022,golami_ab_2022,marek_thermodynamics_2022}. Recently, it has been demonstrated that the formation of these ODMR center is directly related to carbon contamination in hBN.\cite{mendelson_identifying_2021} 

The electronic, optical, and spin properties of the VB$^-$ center has been comprehensively studied in the literature in numerous experimental\cite{gottscholl_initialization_2020,gao_high-contrast_2021,murzakhanov_electronnuclear_2022,haykal_decoherence_2022,liu_coherent_2022,gao_nuclear_2022} and theoretical studies\cite{ivady_ab_2020,sajid_edge_2020,reimers_photoluminescence_2020,barcza_dmrg_2021}. While this center has already been successfully used in various sensing applications\cite{gottscholl_spin_2021,liu_temperature-dependent_2021,tetienne_quantum_2021,healey_quantum_2022,lyu_strain_2022}, single defect measurements have not been demonstrated yet, expectantly due to the center's low photo luminescence (PL) emission rate\cite{ivady_ab_2020}. Furthermore, as reactive nitrogen dangling bonds give rise to the electronic states of the VB$^-$ defect that may be terminated by mobile interstitial and adatoms on the surface, the VB$^-$ center is expectedly chemically unstable in few layer hBN samples. However, antisites and impurities with satisfied covalent bonds may form chemically stable structures with optically addressable electronic states. Spin qubits realized by such defects may remain functional even on the surface and in atomically thin layers. While antisites give rise to only a few structures with no relevant spin properties\cite{WestonPhysRevB2018}, carbon impurities are common in hBN, give rise to numerous complex structures, and form  expectedly the unidentified ODMR centers.

Here, we theoretically study the neutral charge state of the nitrogen and the boron centred symmetric carbon tetramer structures in hBN, i.e.\ (C$_{\text{B}}$)$^3$-C$_{\text{N}}$ and (C$_{\text{N}}$)$^3$-C$_{\text{B}}$ that we dub as C4N and C4B, respectively. The electronic structures of the defects consist of a triplet ground state, an optically allowed triplet excited state, and two singlet states between the triplets. For the C4B defect,  we obtain large inter-system crossing rate from the triplet excited state to the singlet manifold and strain dependent inter-system crossing rate to the ground state. For the C4N defect, the spin selective decay is enabled by out-of-plane distortions. Therefore, strain can be used to engineer the defects' inter-system crossing rates and contrast.  In addition, the carbon tetramers  gives rise to narrow magnetic resonance lines, due to the localization of the spin density on the spinless carbon atoms.  Relying on our results, we propose the C4N and C4B defects in hBN as chemically stable spin qubits for improving sensing in low dimensions.

\begin{figure}[!ht]
\begin{center}
	\includegraphics[width=0.6\columnwidth]{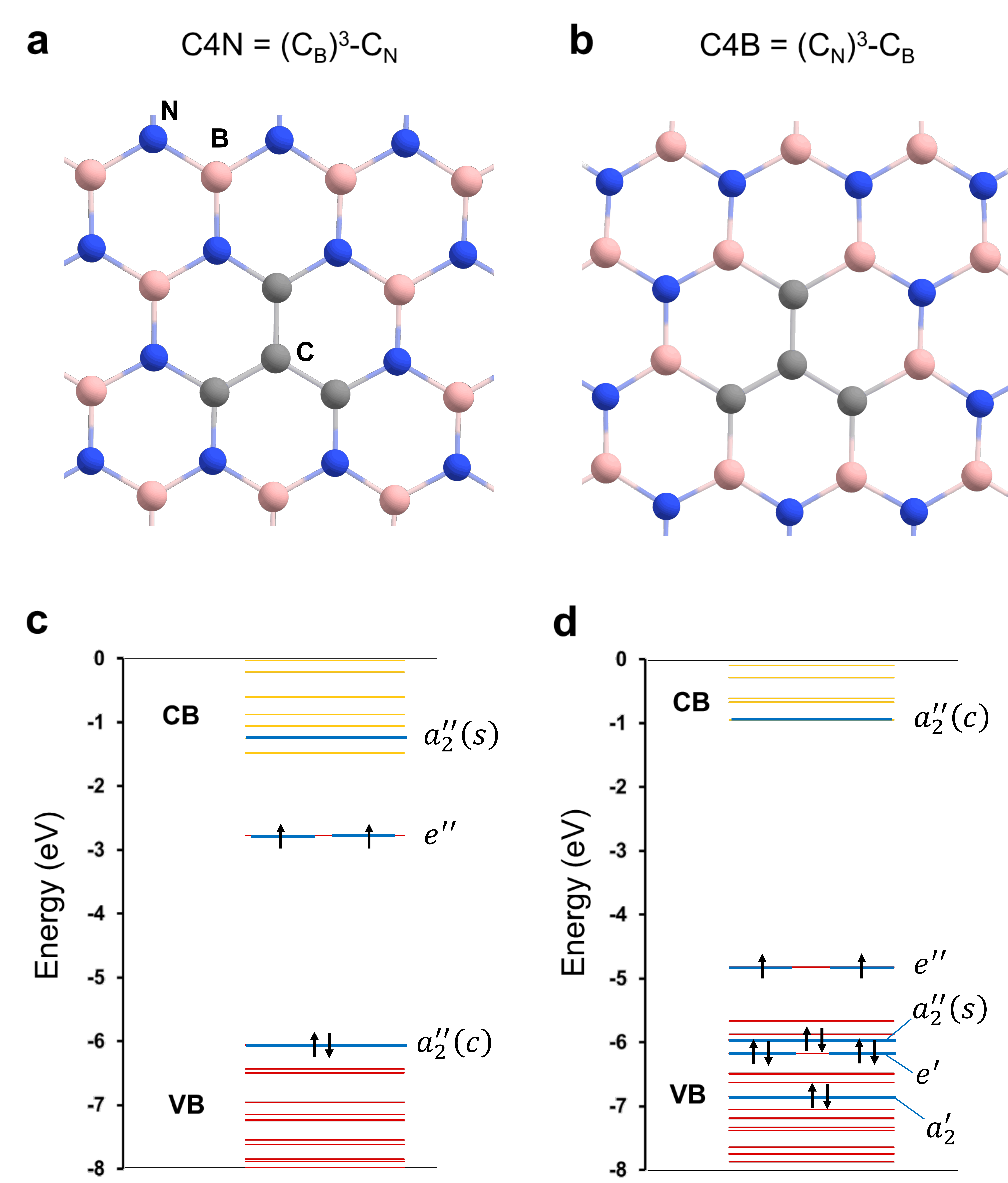}
	\caption{ Atomic and electronic structure of symmetric carbon tetramers in hBN. {\bf a} and {\bf b} show the ground state atomic configuration of the nitrogen site centered C4N defect and the boron site centered C4B defect, respectively. Both defects exhibit D$_{3h}$ point group symmetry.  {\bf c} and  {\bf d} depict the corresponding single particle electronic structures. Yellow, red, and blue lines represents conduction band, valence band, and defect energy levels, respectively. Up and down arrows indicate the occupation of the defect states. The localized defect orbitals are depicted in Supplementary Figure S2.}
	\label{fig:fig1}  
\end{center}
\end{figure}

The thermodynamic properties of carbon complexes have been  studied recently in the literature.\cite{marek_thermodynamics_2022,huang_carbon_2022} Importantly, the symmetric C4N and C4B tetramers, see Fig.~\ref{fig:fig1}a and b, possess a surprisingly low formation energy. This can be explained by the Baird-aromatic stabilization of C4 containing hBN  (i.e. there are two unpaired electrons and altogether $4k$ electrons in the delocalized $\pi$ system of any selected set of rings; $k$ denotes an arbitrary integer), which is similarly favorable to the Hückel aromaticity of pure hBN (i. e. there are $4k+2$ electrons in the delocalized $\pi$ system, all of which are paired). We note here that based on Baird's and Hückel's rules \cite{aromaticity}, the general conclusion can be drawn that even number of carbon atoms can, while odd number of carbon atoms cannot maintain aromaticity, implying larger formation energies in the latter case.

The formation energy of the neutral C4N and C4B defects are 2.5~eV (8.3~eV) and 8.7~eV (2.9~eV) in N-rich (B-rich) growth conditions and further decreases in charged configurations.\cite{marek_thermodynamics_2022} The most relevant charge transition levels are $E(++|+) =  2.38$~eV, $E(+|0)= 3.28$~eV, and $ E(0|-)= 5.59$~eV for the C4N defect and $E(+|0) = 0.48 $~eV, $ E(0|-) = 3.03$~eV, and $ E(-|--) = 3.90$~eV for the C4B defect measured from the valence band maximum.\cite{marek_thermodynamics_2022} The neutral charge state of the C4N (C4B) defect is thus stable in the upper half (lower half) of the band gap, where the Fermi energy is located in N-rich (N-poor) growth conditions.\cite{marek_thermodynamics_2022}

\section*{Results}

The single particle electronic structures of the neutral ground state of the C4N and C4B defects are depicted in Fig.~\ref{fig:fig1}c and d, respectively. In-plane sp2 bonding states of the atoms are fully occupied and most of them fall deep in the valence band. The four $p_z$ orbitals of the carbon atoms form defect states that appear inside the band gap.  In the neutral charge state of the C4N defect, the most relevant single particle defects states are the fully occupied $a_2^{\prime\prime}\left( c \right)$  state, the half occupied $e^{\prime\prime}$  states, and the empty $a_2^{\prime\prime}\left( s \right)$, see Fig.~\ref{fig:fig1}c. As visualized in Supplementary Figure S2, the $a_2^{\prime\prime}$ and $e^{\prime\prime}$ orbitals are primarily located on the central carbon atom and on the three side carbon atoms, respectively. $a_2^{\prime\prime}\left( c \right)$ orbital is strongly localized on the central carbon, while $a_2^{\prime\prime}\left( s \right)$ orbital is slightly delocalized and has significant contributions from the $p_z$ orbitals of the surrounding three boron atoms. The most relevant defects states for the C4B defect  are the fully occupied $e^{\prime}$ and $a_2^{\prime\prime}\left( s \right)$  states, the half occupied $e^{\prime\prime}$  state, and the empty $a_2^{\prime\prime}\left( c \right)$, see Fig.~\ref{fig:fig1}d and Fig.~S2. The occupied defect states of the C4B defect can be found closer to the valence band compared to the case of the C4N defect. For the C4B defect, a fully occupied in-plane bonding $e^{\prime}$ state can be found close to the valance band maximum and it plays an important role in the optical excitation process. Since the double degenerate $e^{\prime\prime}$ state is occupied by two electrons with parallel spin, both defects exhibit a triplet ground state.

\begin{figure}[!ht]
\begin{center}
	\includegraphics[width=0.9\columnwidth]{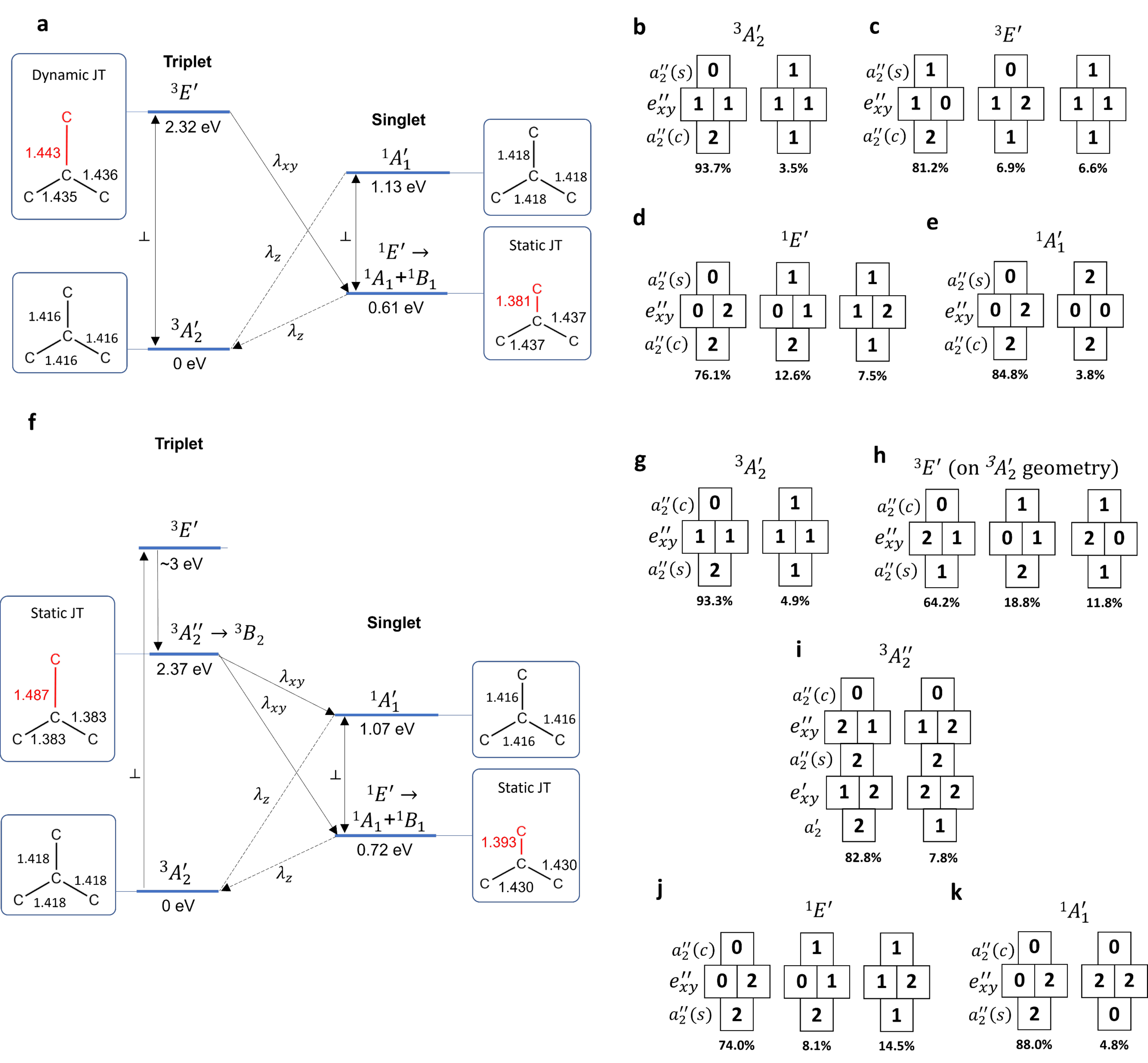}
	\caption{ Many-body electronic structure of the symmetric C4N and C4B defects. {\bf a} Many-body spectrum of the C4N defect separated into singlet and triplet manifolds. Arrows connecting the states indicate possible optical and spin-orbit interaction mediated non-radiative transitions.  {\bf b} -  {\bf e} Slater determinant expression of the many body states of the C4N defect. Squares with numbers indicate the occupancy of each of the localized defect states, while groups of squares represent different Slater determinants. The percentages below the Slater determinants represent the weight of the determinant in the expression. {\bf f} Many-body spectrum of the C4B defect. {\bf g} -  {\bf k} Slater determinant expression of the many body states of the C4B defect.  }
	\label{fig:fig2}  
\end{center}
\end{figure}

The many-body electronic structures of the symmetric carbon tetramers are schematically visualized in Fig.~\ref{fig:fig2}, where the energy gaps are obtained with an anticipated error margin of  $\pm0.15$~eV\cite{NEVPT2_error} on CASSCF-NEVPT2 level of theory\cite{nevpt2}. As can be seen, four (five) states can be found in the low energy part of the electronic structure of the C4N (C4B) defect. The ground states of both defects can be described to a large degree ($>\!93$\%) by the single Slater determinant obtained in the single particle picture in DFT, see Fig.~\ref{fig:fig2}c-d. For C4N, a $^3E^{\prime}$ triplet state can be found 2.32~eV above the $^3A_2^{\prime}$ ground state. (The corresponding zero phonon photoluminescence (ZPL) energy, which  is obtained by adding the zero-point energy contribution of the local vibrational modes to the 2.32~eV adiabatic energy difference, was found to be 2.18~eV.) The $^3E^{\prime}$ optically excited state is largely described by the determinant corresponding to the $e_x^{\prime\prime} \rightarrow a_2^{\prime\prime} \! (s)$ transition, however, other determinants of single excitation mix with the leading term as depicted in Fig.~\ref{fig:fig2}\textbf{c}. Transition between the triplets is enabled by  parallel to $c$ polarized photon absorption and emission. The $^3E^{\prime}$ state is Jahn-Teller (JT) unstable and the optimized structure is slight distorted, see Fig.~\ref{fig:fig2}\textbf{a}. Due to the small JT distortion, we expect a dynamic Jahn-Teller effect, where the vibronic $^3\widetilde{E}^{\prime}$ state exhibits effective high symmetry. (Here, we note that an alternative, out-of-plane distorted triplet excited state geometry was also observed in some calculations, see Supplementary Information, Figure~S5. This effect is however peculiar to single, separated hBN layers.) In between the triplet states, there are two singlet excited states, a $^1E^{\prime}$ state and a $^1A_1^{\prime}$ state 0.61~eV and 1.13~eV above the ground state energy level, respectively. 

For the C4B defect, we obtain a similar many-body electronic structure as for the C4N defect; however, with an additional dark triplet excited state in-between the ground state and the $^3E^{\prime}$ triplet excited state, see Fig.~\ref{fig:fig2}a and f. The lower lying $^3A_2^{\prime\prime}$ excited state and the optically excited $^3E^{\prime}$ state  is expected to be found  {2.37}~eV and $\sim${3.0}~eV above the ground state. (Note that the latter value is an estimation based on the vertical excitation energy of 3.7 eV at the $^3A_2^{\prime}$ geometry and approximate structural relaxation energy as the relaxation to $^3E^{\prime}$ geometry could not be performed due to the delocalization of the characteristic orbitals.) These states are described to a large degree by an $e^{\prime} \rightarrow e^{\prime\prime}$ transition and $a_2^{\prime} \rightarrow e^{\prime\prime}$ transition, respectively, see Fig.~\ref{fig:fig2}h and i. Both states are JT unstable and the point group symmetry of the optimized excited state structure reduces to C$_{2\text{v}}$. The relaxed lowest energy triplet excited state belongs to the $B_2$ irreducible representation of C$_{2\text{v}}$. In D$_{3\text{h}}$ symmetry, no optical transition is possible between the ground and the $^3A_2^{\prime\prime}$ excited state. This property is largely preserved even in the JT distorted $^3B_2^{\prime\prime}$ state. Transition to the higher lying $^3E^{\prime}$ state is possible, similarly to the C4N defect; however, due to the proximity of the $^3B_2^{\prime\prime}$ state a rapid non-radiative decay to this lower lying triplet state is expected.  Therefore, the C4B defect can be optically excited, but no PL emission is possible from this defect in the visible range. In addition to the triplets, we find two singlets, a $^1E^{\prime}$ state and a $^1A_1^{\prime}$ state 0.72~eV and 1.07~eV above the ground state, respectively. 

Slater determinant expansion of the singlet states are also provided in Fig.~\ref{fig:fig2}\textbf{b}-\textbf{e} and Fig.~\ref{fig:fig2}\textbf{g}-\textbf{k} for the C4N and C4B defects. The most relevant $^1E^{\prime}$ and the $^1A_1^{\prime}$ states are the singlets that correspond to the ground state occupancy of the single particle orbitals. These singlets mix with other excited determinants to a similar degree as the optically excited states. The $^1E^{\prime}$ state is a Jahn-Teller unstable states and it goes through a severe structure distortion. In the optimized configuration one of the C-C bonds shortens that splits the $^1E^{\prime}$ state into a $^1A_1$ and a $^1B_1$ states in C$_{2v}$ symmetry, see Fig.~\ref{fig:fig2}a and f. 

\begin{figure}[!ht]
\begin{center}
	\includegraphics[width=0.80\columnwidth]{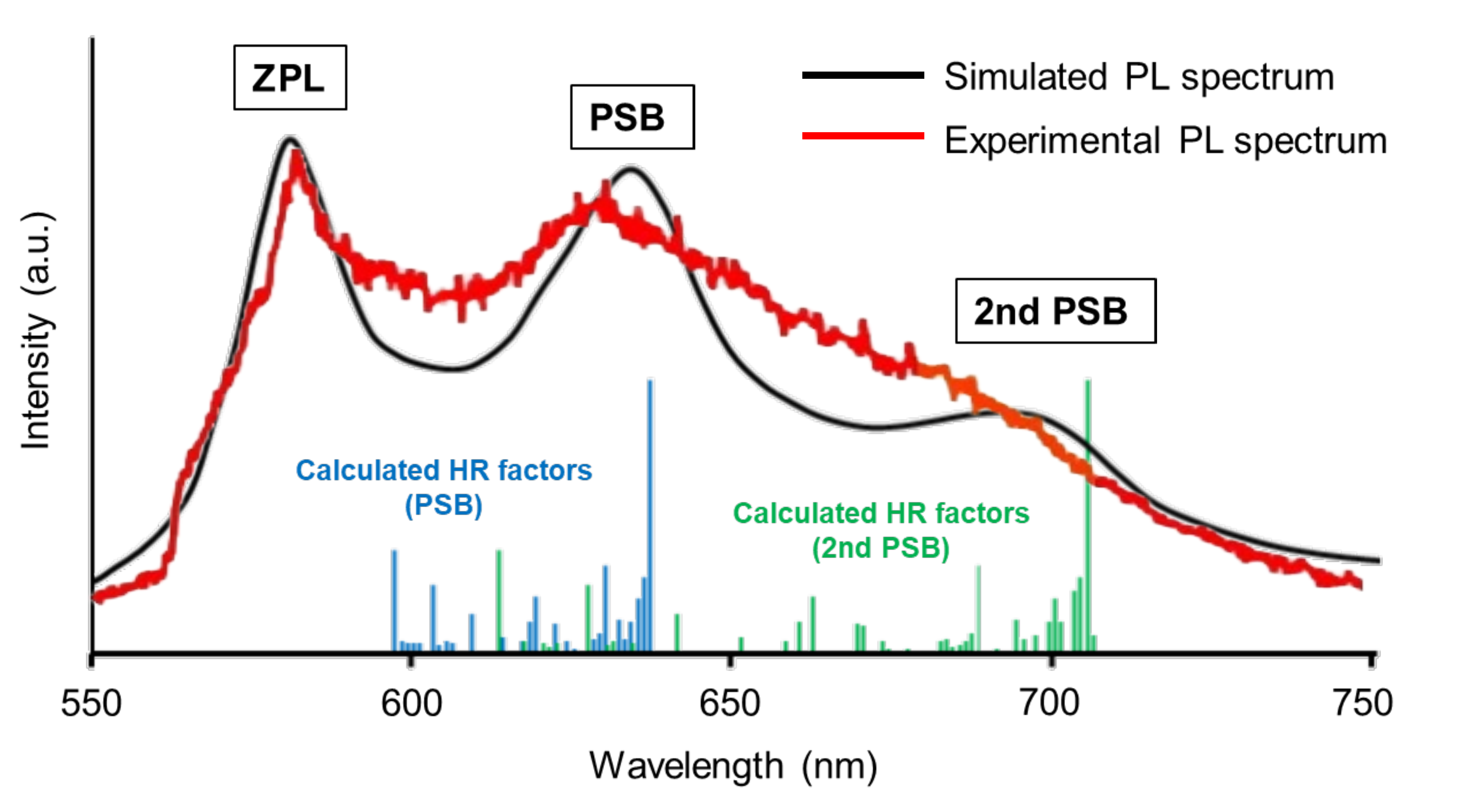}
	\caption{ Photoluminescence spectrum of the symmetric C4N defect in hBN compared with an experimental signal\cite{mendelson_identifying_2021} of the unidentified carbon related spin qubit in hBN. For better comparison the  theoretical ZPL emission is aligned  with the first maximum of the experimental spectrum. The theoretical spectrum is broadened to mimic temperature and inhomogeneous effects. Blue and green impulses show the partial Huang Rhys factors of one and two photon processes. }
	\label{fig:pl}  
\end{center}
\end{figure}

For the radiative lifetime of the triplet excited states of the C4N defect, we obtain 80.5~ns at 0~K using 2.13 refractive index for a 590~nm photon. From the computed dipole moment of the ${}^{3}E^{\prime} \rightarrow {}^{3}A^{\prime}_2$ transition and harmonic vibrational analysis of the two individual electronic states, we calculate the phonon side band (PSB) of the C4N defect, see Fig.~\ref{fig:pl}.  We obtain 1.8 for the Huang Rhys factor (HR factor) that correspond to Debye Waller factors (DW factors) of 0.165. The relatively high DW factor for C4N defect is due to the high symmetry and small distortion of the excited state geometry. In addition, in $D_{3h}$ symmetry we expect weak coupling to electric field for the C4N structure.\cite{zhigulin_stark_2022}

Partial HR factors of the vibrational modes indicate that the PL transitions couple strongest to in-plane carbon-carbon bond stretching modes that drive the JT distorted excited state into the symmetric ground state configuration (see section S3.3 of the Supplementary Information for the visualization of the vibrations). The energy of both modes is 0.196~eV (1583 cm$^{-1}$), which corresponds to the generally observed range of aromatic C-C stretching in infrared spectroscopy.\cite{aromatic_stretching} On the other hand, additional lower energy modes also couple to the defect that broaden and shift the maximum of the resonance peaks of the calculated PL spectrum. The phonon side-band of the PL emission of the C4N defect is compared with the experimental PL spectrum\cite{mendelson_identifying_2021,stern_room-temperature_2022} of the unidentified carbon related quantum bit, see Figure~\ref{fig:pl}. The theoretical and experimental spectra agree very well.

\begin{table}
\begin{center}
\caption{\label{tab:LSdata} Spin-orbit coupling matrix elements (SOCMEs) as obtained on CASSCF-NEVPT2 level of theory for 0~K optimised and distorted atomic configurations. The distortion is quantified by the dihedral angle ($\phi$) corresponding to the position of the 4 carbon atoms in the ground state. See section S3.2 of SI for the visualization of distorted geometries. All values are in GHz. We note that the matrix element depends on the geometry. Herein, the average of initial-state and final-state SOCMEs is given. When a zero and a non-zero SOCMEs are obtained for the involved inital and final electronic states, the non-zero value is provided.}
 \begin{tabular}{|c|ccc|| c|c| }
 \hline
 \multicolumn{4} {|c||}{ C4N } & \multicolumn{2} {c|}{ C4B } \\ \hline
 \multirow{3}{*}{  Transition} & \multicolumn{3}{c||}{matrix element } & \multirow{3}{*}{  Transition} &  matrix \\ \cline{2-4}
 &  0~K static & Buckling  & Buckling & & element \\ \cline{6-6}
 &  ($\phi$=0°) &  ($\phi$=2.7°) & ($\phi$=6.9°) & &  0~K static \\ \hline
 $^3E^{\prime}(m_s = \pm1) \rightarrow {^1}A_1^{\prime}$ & 0 & 16.03 & 31.95 & $^3A_2^{\prime\prime}(m_s = \pm1) \rightarrow {^1}A_1^{\prime} $ &  27.08 \\
 $^3E^{\prime} (m_s = 0) \rightarrow {^1}A_1^{\prime}$ & 0.03  & 0.03 & 0.18 &  $^3A_2^{\prime\prime} (m_s = 0) \rightarrow {^1}A_1^{\prime}$ & 0  \\
 $^3E^{\prime} (m_s = \pm1) \rightarrow {^1}E^{\prime}$ & 0 &  7.67 & 13.26 & $^3A_2^{\prime\prime} (m_s = \pm1) \rightarrow {^1}E^{\prime}$ & 48.81  \\
 $^3E^{\prime} (m_s = 0) \rightarrow {^1}E^{\prime}$ & 0.45 & 4.65 & 14.94 & $^3A_2^{\prime\prime} (m_s = 0) \rightarrow {^1}E^{\prime}$ & 0  \\
 $^1A_1^{\prime} \rightarrow {^3}A_2^{\prime} (m_s = \pm1)$ & 0 & 0.26 & 0.15 & $^1A_1^{\prime} \rightarrow {^3}A_2^{\prime} (m_s = \pm1)$ & 0  \\
 $^1A_1^{\prime}\rightarrow {^3}A_2^{\prime} (m_s = 0)$ & 2.76 & 2.61 &  11.25 & $^1A_1^{\prime}\rightarrow {^3}A_2^{\prime} (m_s = 0)$ & 2.71 \\
 $^1E^{\prime} \rightarrow {^3}A_2^{\prime} (m_s = \pm1)$ & 0 & 6.41 &  9.53 & $^1E^{\prime} \rightarrow {^3}A_2^{\prime} (m_s = \pm1)$ & 0 \\
 $^1E^{\prime} \rightarrow {^3}A_2^{\prime} (m_s = 0)$ & 0.27  & 0.24 & 1.47 & $^1E^{\prime} \rightarrow {^3}A_2^{\prime} (m_s = 0)$ & 0.06 \\ \hline
 \end{tabular}
\end{center}
\end{table}

Point defect quantum bits are a special type of color centers that exhibit high spin ground state and spin state dependent optical emission through spin selective non-radiative decay processes from the optically excited state to the ground state. As  carbon tetramers possess high spin ground state and singlet shelving states between the triplet excited and ground states, they may implement optically addressable spin quantum bits in hBN that we investigate in the following. 

In order to obtain an efficient spin state dependent non-radiative decay channel through the singlets, the triplet and the singlet manifolds should be coupled by strong spin-orbit coupling (SOC) matrix elements. For the C4B defect we obtain 48.81~GHz (27.08~GHz) spin-orbit coupling matrix elements between the $^3A_2^{\prime\prime}$($m_S = \pm1$) and the ${^1}E^{\prime}$ (${^1}A_1^{\prime}$) states, see Table~\ref{tab:LSdata}. Spin-orbit interaction thus gives rise to a decay channel with approximately 0.07~MHz decay rate for the dark excited triplet of the C4B defect. The spin-orbit coupling matrix elements between the $^{1}E^{\prime}$ state and the ground state are weaker, enabled mostly by the Jahn-Teller distortion of the low lying singlet state. Consequently, the $^{1}E^{\prime}$ state is long-lived compared to the triplet excited state. These results indicate that the C4B defect can be spin polarized through an optical excitation to the ${^3}E^{\prime}$ state and subsequent non-radiative and spin selective decay through the ${^3}A_2^{\prime\prime}$, ${^1}A_1^{\prime}$, and ${^1}E^{\prime}$ states. Due to this behaviour, the C4B defect may observed in electron spin resonance (ESR) measurement in low concentrations under $\sim$3.0~eV excitation. In addition, due to the strict spin selectivity of the non-radiative decay from the ${^3}A_2^{\prime\prime}$ excited state and the expectedly long lifetime of the ${^3}A_2^{\prime\prime}$($m_S = 0$) state, the defect may be suitable for photoelectron detected magnetic resonance (PDMR) read-out of the spin states.

The C4N defect's lowest energy triplet excited state has a different symmetry than the excited state of the C4B defect, therefore, the spin-orbit coupling between the triplets and the singlets is forbidden in first order approximation in D$_{3h}$ symmetry. Indeed, considering the 0~K static atomic structure and corresponding many-particle electronic states, we obtain either zero or small spin-orbit coupling matrix elements between the states on CASSCF-NEVPT2 level of theory, see Table~\ref{tab:LSdata}, left column. The non-zero elements are due to the JT effect that give rise to weak couplings between states. On the other hand, strain and out-of-plane distortions may break the symmetry that could give rise to a sizable increases of the spin-orbit coupling matrix elements for C4N. For example, in-plane compression of hBN flakes causes the sample to buckle, which can be quantified by the dihedral angle given by the position of the 4 carbon atoms (denoted as $\phi$ in Table~\ref{tab:LSdata}). $\phi$ equals to 0° in the equilibrium geometry as all C atoms are located in one plane, but it shows an increasing deviation from 0° when enforcing the side of the flake - i.e.\ the outside B and N atoms - to remain in fixed position closer and closer to the center (see section S3.2 of Supplementary Information for representative geometries).  Buckling of hBN, which is commonly observed in experiments \cite{stern_room-temperature_2022}, can give rise to spin-orbit matrix elements in the 10-30~GHz range that are comparable with NV center's matrix elements.\cite{thiering_ab_2017} The increased SOC matrix elements give rise to spin dependent non-radiative dacay channels that may facilitate optical read-out of the spin state.

As a demonstrative example, we computed the photoluminescence and inter-system crossing rates between the electronic states of the buckled flake at $\phi$=6.9° - see section S3.4 of the Supplementary Information. The obtained data indicate polarizability in $m_S = \pm 1 $ state. These results demonstrate that the C4N defect can implement an optically addressable spin qubit, whose spin dependent optical signal is enabled mostly by local strain.

Apart from strain induced geometry distortions, the out-of-plane vibration of the C4 center may also give rise to significant SOC matrix elements and - consequently - intersystem crossing rates in the case of C4N. Namely, even though the matrix elements are zero (or close to zero) in equilibrium geometry, it does not hold true for non-equilibrium geometries where the system spends a significant amount of time due to its constant vibration (See Table S6 for SOC matrix elements computed at geometries displaced along the out-of-plane normal mode). Thus, a Herzberg-Teller transition\cite{Herzberg-Teller} is possible, the rate of which strongly depends on the occupancy of vibrational levels (i. e. the temperature).

\begin{table}
\begin{center}
\caption{\label{tab:HFdata} Hyperfine tensors of C4N and C4B defect for the strongest coupled nuclear spins. Location of the considered nuclear spins are visualized in Fig.~\ref{fig:spin}.  In the calculations, we use $^{13}$C, $^{11}$B, and $^{14}$N isotopes. The table provides the hyperfine tensor for one of the symmetrically equivalent positions and N gives  the number of equivalent positions in the lattice. The complete table of hyperfine tensors can be found in Table~S3.6 of the Supplementary Information. All values are in MHz. }
 \begin{tabular}{|c|c|cccc|| c|c|cccc| }
 \hline
 \multicolumn{6}{|c||}{C4N}  &   \multicolumn{6}{c|}{C4B} \\ \hline
 site & N & $A_{xx}$ & $A_{yy}$ & $A_{zz} = A_z$ & $A_{xy}$ & site & N & $A_{xx}$ & $A_{yy}$ & $A_{zz} = A_z$ & $A_{xy}$  \\ \hline
C$_c$ & 1 & -33.4 & -33.4 & -57.6 & 0.0 & C$_c$ & 1 & -22.9 & -22.9 & -48.0 & 0.0 \\ 
C$_s$ & 3 & 8.6 &  8.8 &  95.0 &  0.0 & C$_s$ & 3 & -2.6 & -2.6 & 68.9 & 0.0  \\
N$_1$ & 6 & -3.2 &  -3.3 &  0.3  & 0.1 & B$_1$ & 6 & -10.6 & -8.5 & -6.8 & -0.2 \\
B$_2$ & 3 & -0.6 &  -0.3  &  1.7 &  0.0 & N$_2$ & 3 & -0.5 & -0.3 & 1.5 & 0.0 \\
B$_3$  & 6 & 0.2 &  -0.9  &  1.5  &  0.3 & N$_3$ & 6 & -0.1 & -0.4 & 1.1 & 0.0  \\ \hline
B$_{A/B}$ & 2 &  -0.3 &  -0.3  &  0.6 & 0.0 & B$_{A/B}$ & 6 & -0.2 & -0.2 & 0.6 & 0.0 \\ \hline
 \end{tabular}
\end{center}
\end{table}

\begin{figure}[!ht]
\begin{center}
	\includegraphics[width=0.70\columnwidth]{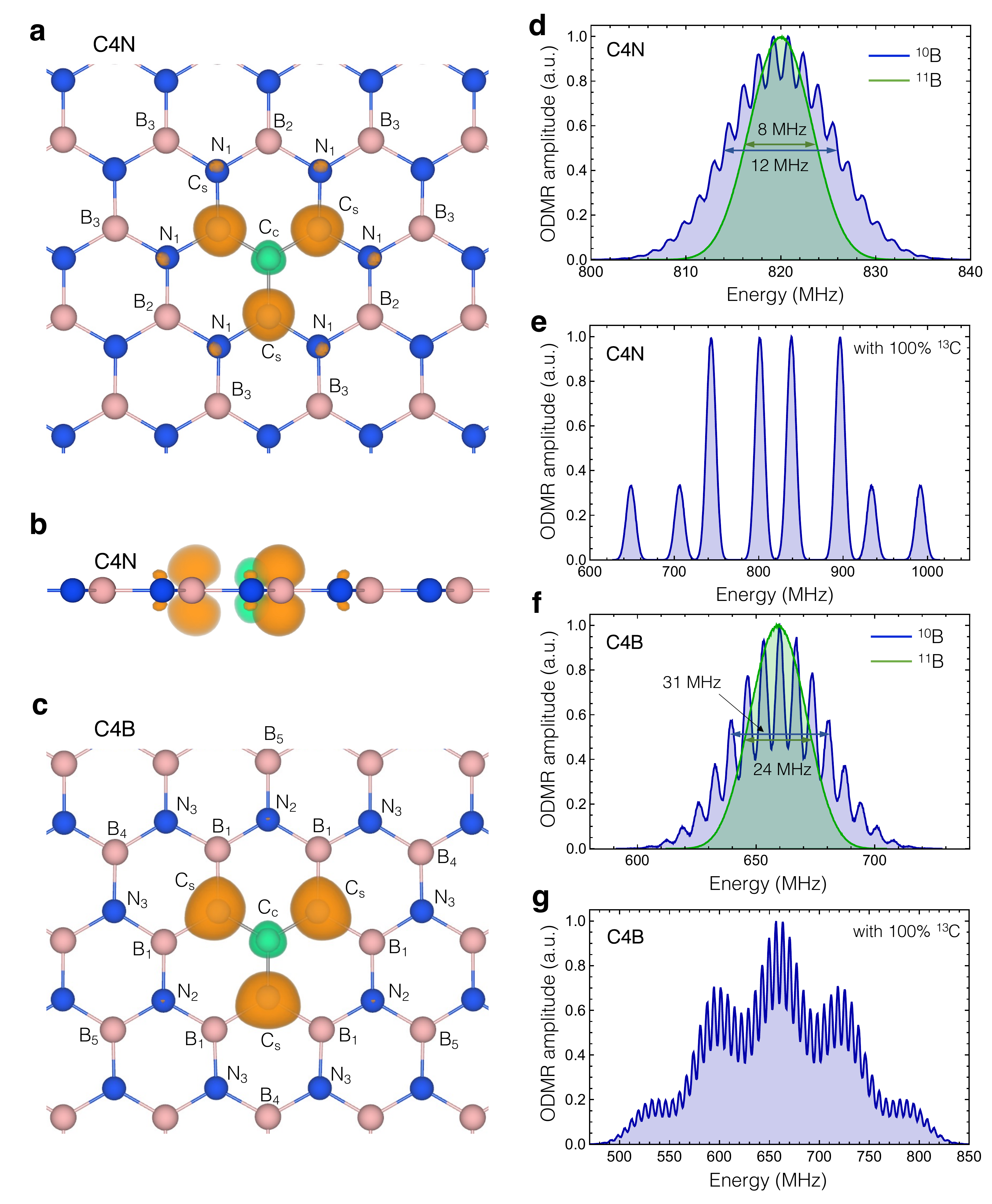}
	\caption{ Spin densities and spin resonance signals of symmetric carbon tetramers. \textbf{a} and \textbf{b} depict the spin density of the C4N defect from top and side views. \textbf{c} depicts the spin density of the C4B defect from the top view. \textbf{d} Electron spin resonance signal of the C4N defect at $B = 0$ with no carbon nuclear spins in $^{10}$B (blue) and $^{11}$B (green) containing samples. \textbf{e} Electron spin resonance signal of the C4N defect when it includes four $^{13}$C nuclear spins. \textbf{f} Simulated electron spin resonance signal of the C4B defect with no carbon nuclear spins in $^{10}$B and $^{11}$B containing samples. \textbf{e} Spin resonance signal of the C4B defect including four $^{13}$C nuclear spins.}
	\label{fig:spin}  
\end{center}
\end{figure}

Finally, we present our result on the spin properties of the C4N and C4B defects. For the zero-field splitting (ZFS) parameter $D$ of the C4N defect we obtain 820~MHz and 840~MHz with CASSCF-NEVPT2 and periodic hybrid-DFT with spin contamination error correction\cite{biktagirov_spin_2020}, respectively. The difference of the values indicate the error margin of our ZFS calculations. For the C4B defect we obtain a slightly reduced $D = 660$~MHz ZFS value on CASSCF-NEVPT2 level of theory.  Due to the D$_{3h}$ symmetry, the quantization axis of the defects is parallel to the $c$-axis and no $E$ splitting is observed in unstrained configurations. The spin density of the defects and the locations of the most relevant nuclear spins  are depicted in Fig.~\ref{fig:spin}a-c. The corresponding hyperfine coupling parameters, obtained with periodic hybrid-DFT in a bulk model, are provided in Table~\ref{tab:HFdata} and in section S3.6 of the Supplementary Information. Most notably, the spin density localizes mainly on the carbon atoms and only secondary localization can be found on the first neighbor nitrogen atoms. Accordingly, the carbon hyperfine parameters an order of magnitude larger than the rest of the coupling parameters. Using the theoretical spin coupling parameters, the predicted electron spin resonance (ESR) spectra at zero magnetic field for different isotope abundances are depicted in Fig.~\ref{fig:spin}d-g. In natural abundance, $^{13}$C nuclear spin can be found only with 1.07\% probability, thus in most configurations no carbon spins are included in the structures. When completely ignoring carbon nuclear spins, we obtain a narrow homogeneous ESR signal, where the line width is determined by the boron hyperfine coupling tensor and the boron isotope abundance. The full widths of the resonance peaks at half maximum are 12~MHz (8~MHz) and 31~MHz (24~MHz) for the C4N and the C4B defects in ${^{10}}$BN (${^{11}}$BN) sample. The narrow ESR line width may make the C4N and C4B defects attractive candidates for sensing. When carbon nuclear spins are included with 100\% $^{13}C$ abundance, we observe a characteristic 8 peak hyperfine structure with 95.0 and 57.6 MHz splittings for the C4N defect, see Fig.~\ref{fig:spin}e, and a characteristic 5 broad peak structure with $\sim$60~MHz splitting. $^{13}$C enriched carbon contamination and observation of the carbon related hypefine structure can be used to unambiguously identify the C4N and C4B defects.

\section*{Discussions}

\bigbreak
\noindent \textit{Sensing applications}

Recently, few nanometers thick sensing foils has been developed by using spin qubit-containing hBN sheets.\cite{tetienne_quantum_2021,healey_quantum_2022,kumar_magnetic_2022} In order to maximally utilize the layered structure of hBN and to further boost sensitivity and spatial resolution in such experiments, single sheet hBN flakes with stable point defect spin qubits are needed. We propose here the charge neutral symmetric carbon tetramers for implementing such chemically stable quantum bits in hBN. Due to the bond-formations, the carbon tetramers can give rise to atomic thin sensing foils that may be stable even at ambient conditions. Furthermore, the narrow electron spin resonance linewidth of the nitrogen centered carbon tetramer may lead to high sensitivity.

Robust spin qubits could also be highly beneficial for hBN-based dynamic nuclear polarization.\cite{gao_nuclear_2022} We show here that carbon tetramers can be spin polarized by optical illumination, leading to a spin polarization sources that may couple to nuclear spins outside the hBN host. Since the symmetric carbon tetramers can be stable at the surface in close proximity to other molecules, e.g. the NMR agent tetramethylsilane, efficient polarization transfer could be achieved between the quantum defects and the molecules on the surface. High-temperature and low-magnetic nuclear hyperpolarization mechanism are long-time sought for for boosting the sensitivity of conventional NMR and MRI applications.

Stable spin qubits are key components of both of the above-mentioned applications. Since, the symmetric carbon tetramers improve on the state of the art in this respect, they could lead to advances in nanoscale sensing.

\bigbreak
\noindent \textit{Fabrication}

Previous computational studies\cite{marek_thermodynamics_2022} have demonstrated that carbon tetramers exhibit low formation energy in N-rich samples that may imply the formation of these complex defects in observable concentrations ($\sim 10^{14}$ cm$^{-3}$) in hBN. After growth treatments, however, may give rise to concentrations well exceeding the thermal equilibrium values and enable on-demand fabrication in few layer samples. In this respect, carbon implantation maybe of high importance that can create carbon related point defect quantum bits in hBN.\cite{mendelson_identifying_2021} Furthermore, scanning transmission electron microscope (STEM) is also of high potential for sub-nanometer precision creation of defects.\cite{park_atomically_2021} STEM mapping combined with a subsequent annealing steps allows the creation of the carbon complexes in single layer hBN samples.\cite{park_atomically_2021} These processes open up new directions for tailored fabrication of carbon clusters, including the symmetric C4N defect, in hBN.

\bigbreak
\noindent \textit{Summary}

In this paper we comprehensively studied the neutral nitrogen and boron site centered symmetric carbon tetramers in hBN. By using complementary first principles methods, we predicted the electronic structure as well as the optical and spin properties of the defects. We showed that symmetric carbon tetrameters exhibit high spin ground state, optical transition in the visible range, singlet shelving states, and strain dependent non-radiative decay channels. We concluded that the C4N and C4B defects can implement chemically stable spin qubit that may remain functional on the surface even at ambient conditions, in contrast to the established vacancy related quantum defects. Furthermore, strain can be used to tailor the spin contrast of the defects, which is a new feature for spin quantum bits in hBN.

\section*{Methods}

In this study, we carried out first principles calculations of both periodic, supercell model and molecule model of the symmetric C4 defects in hBN. The two approaches allowed independent computational studies on different levels of theory, ensuring the reliability of our results.
Below we briefly summarize the computational protocols; the detailed description of the methodology (including sample input files for the sake of reproducibility) can be found in the Supplementary Information.

\begin{figure}[!ht]
\begin{center}
	\includegraphics[width=0.80\columnwidth]{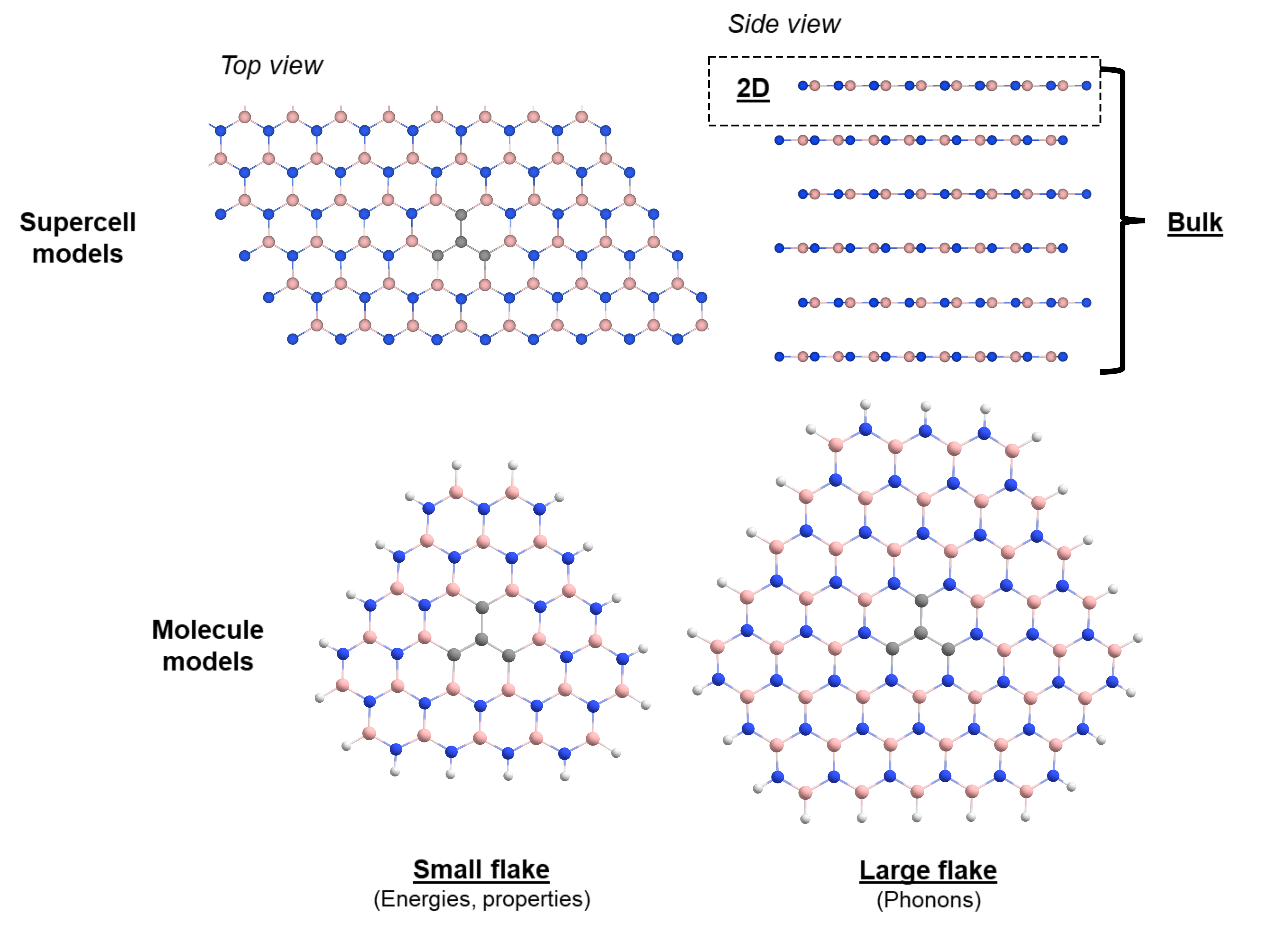}
	\caption{ Illustration of supercell and molecular models used in the present study for the case of C4N defect.}
	\label{fig:models}  
\end{center}
\end{figure}

\bigbreak
\noindent \textit{  Computations with supercell models}

Kohn-Sham density functional theory was employed to study the periodic models of carbon tetramer defects with VASP package\cite{VASP}. Here, we considered a plane wave basis set of 450 eV, PAW\cite{PAW} core potentials, and HSE06 hybrid exchange-correlation functional\cite{HSE03} with 0.32 exact exchange fraction\cite{WestonPhysRevB2018}. The defects were modeled in 162-atom (monolayer) and 768-atom (bulk) supercells (Fig. \ref{fig:models}, top). To account for van der Waals interaction, we included the D3 correction by Grimme et al\cite{DFT-D3-Grimme}. The ZPL energies were calculated from the energy difference between ground and excited states, where the excited states were obtained using spin-conserved constrained DFT method\cite{dSCF} without enforcing symmetry restrictions.

\bigbreak
\noindent \textit{ Computations with molecule models }

As a reasonable compromise between cost and accuracy, two hBN flakes of different size were used to investigate the C4 defects as a finite molecule. Electronic energies and properties (e.g. transition dipole moments) were computed on a smaller model (Fig. \ref{fig:models}, bottom left), while phononic effects were calculated on a larger system (Fig. \ref{fig:models}, bottom right). Test calculations justifying the choice of flake size can be found in the Supplementary Information. All calculations were carried out using the ORCA 5.0.3. program\cite{neese2022software}.

 Geometry optimizations and vibrational analyses were performed using density functional theory, at PBE0/cc-pVDZ level\cite{pbe0func,cc-pVDZ} with D3(BJ) dispersion correction\cite{d3bj}. In the case of excited states, time-dependent density functional theory (TD-DFT)\cite{td-dft} was requested with 10 roots. Singlet excited states were generated from the triplet ground state by spin-flip\cite{spin-flip}.

Single-point energies and electronic properties were determined at CASSCF/cc-pVTZ level of theory\cite{casscf,cc-pVDZ} while dynamic correlation energy was taken into account by second-order N-electron valence perturbation theory (NEVPT2)\cite{nevpt2}. 
The active orbital space for CASSCF was constructed based on both time-dependent density functional theory results obtained by ORCA and density matrix renormalization group (DMRG)\cite{reiher-dmrg,legeza-dmrg} calculations using the Budapest-DMRG package\cite{budapest_qcdmrg}.

\section*{Acknowledgments} 

This research was  supported by the National Research, Development, and Innovation Office of Hungary  within the Quantum Information National Laboratory of Hungary (Grant No. 2022-2.1.1-NL-2022-00004) and within grants FK 135496 and FK 145395.
V.I. also appreciates support from the Knut and Alice Wallenberg Foundation through WBSQD2 project (Grant No.\ 2018.0071). O.L. was supported by the Center for Scalable and Predictive methods for Excitation and Correlated phenomena (SPEC), funded as part of the Computational Chemical Sciences Program by the U.S. Department of Energy (DOE), Office of Science, Office of Basic Energy Sciences, Division of Chemical Sciences, Geosciences, and Biosciences at Pacific Northwest National Laboratory.
The calculations were performed on resources provided by the Swedish National Infrastructure for Computing (SNIC) at the National Supercomputer Centre (NSC).
We acknowledge KIF\"U for awarding us access to computational resource based in Hungary.
Z.B. and T.S. would like to thank the University of Alabama and the Office of Information Technology for providing high performance computing resources and support that have contributed to these research results. 

\section*{Data availability}

The data that support the findings of this study are available from the authors upon reasonable request.

\section*{Author contributions}

Z.B., R.B, and A.G. carried out the first principles calculations, O.L. developed the DMRG program package. V.I., Z.B., and G.B. wrote the manuscript with inputs from all coauthors. The work was supervised by V.I. and G.B. 

\section*{Competing interests}

The authors declare no competing interests.

\section*{References}
\bigbreak

\end{document}


\maketitle
\date{
\noindent
$^1$ Strongly Correlated Systems Lend\"{u}let Research Group, Wigner Research Centre for Physics, PO Box 49, H-1525, Budapest, Hungary\\
$^2$ MTA-ELTE Lend\"{u}let "Momentum" NewQubit Research Group, P\'{a}zm\'{a}ny P\'eter, S\'et\'{a}ny 1/A, 1117 Budapest, Hungary\\
$^3$ Department of Chemical and Biological Engineering, The University of Alabama, Tuscaloosa, Alabama 35487, United States\\
$^4$ Department of Physics of Complex Systems, E\"otv\"os Loránd University, Egyetem t\'er 1-3, H-1053 Budapest, Hungary\\
$^5$ Department of Physics, Chemistry and Biology, Link\"oping University, SE-581 83 Link\"oping, Sweden\\
$^*$email: barcza.gergely@wigner.hu, ivady.viktor@ttk.elte.hu
}


\tableofcontents

\section{Detailed description of the computational methodology (molecular models)} 

\subsection{Construction of molecular (flake) models}

\begin{figure}[H]
\begin{center}
	\includegraphics[width=0.90\textwidth]{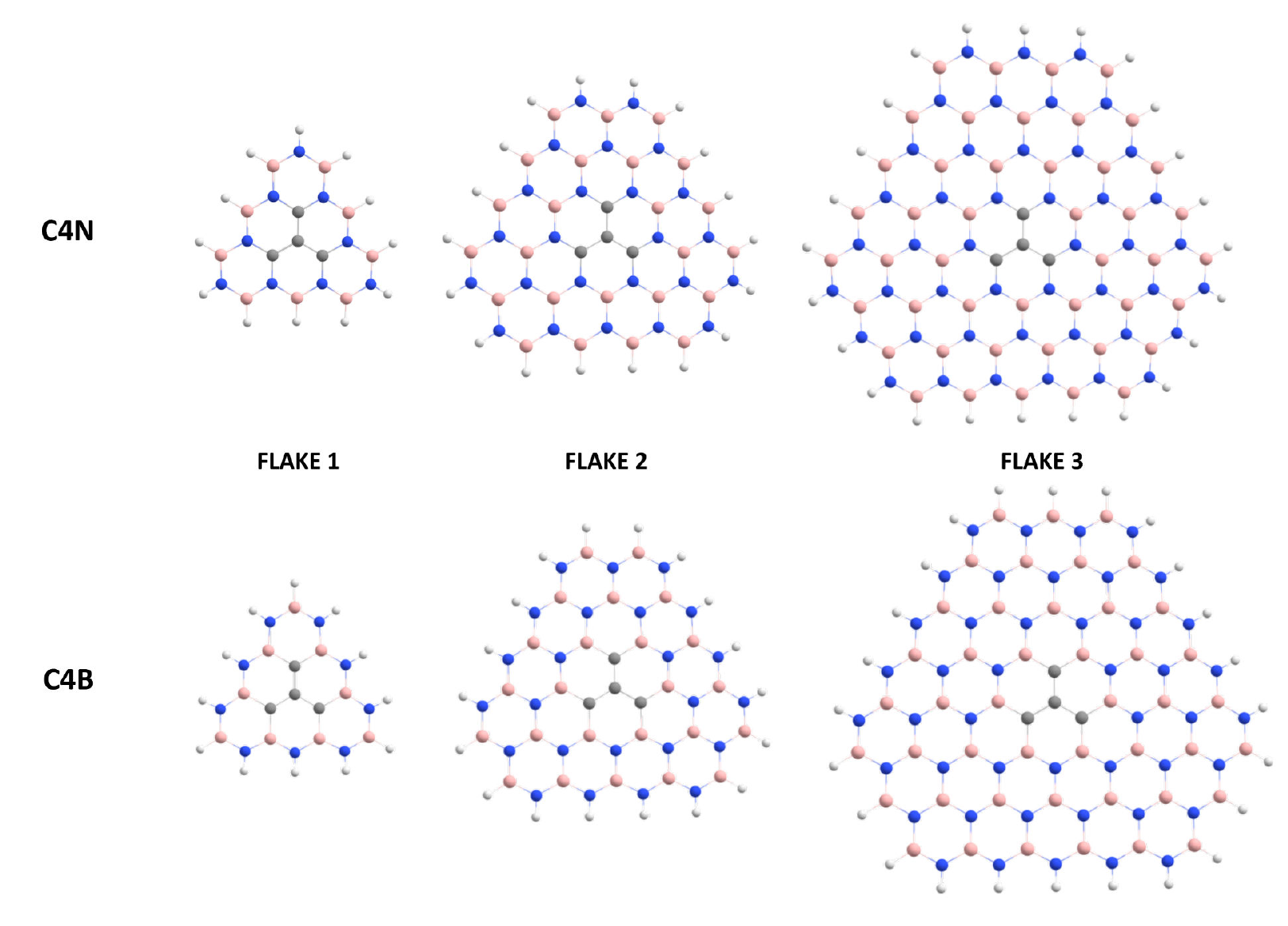}
	\caption{ Molecular structure of the considered flake models. In the ball-and-stick representation generated by Chemcraft program package\cite{chemcraft}, hydrogens, borons, carbons and nitrogens are colored white, pink, gray, and blue, respectively.  We note that Flake 2 and Flake 3 are referred to as "small flake" and "large flake", respectively, in the article.}
	\label{fig:flakes}  
\end{center}
\end{figure}
In order to study the C4 defects based on finite molecular models, we designed three flake structures. As shown in Figure \ref{fig:flakes}, we initially formulated a minimal model (FLAKE 1) and systematically increased the number of pristine boron nitride hexagons around the defect, generating FLAKE 2 and FLAKE 3. As the first step of our investigations, we optimized the geometry of these models in their ground triplet state ($^{3}A_2^\prime$).

\subsection{Ground state orbital diagrams, shape of frontier (C4 centered) molecular orbitals}

\begin{figure}[H]
\begin{center}
	\includegraphics[width=\columnwidth]{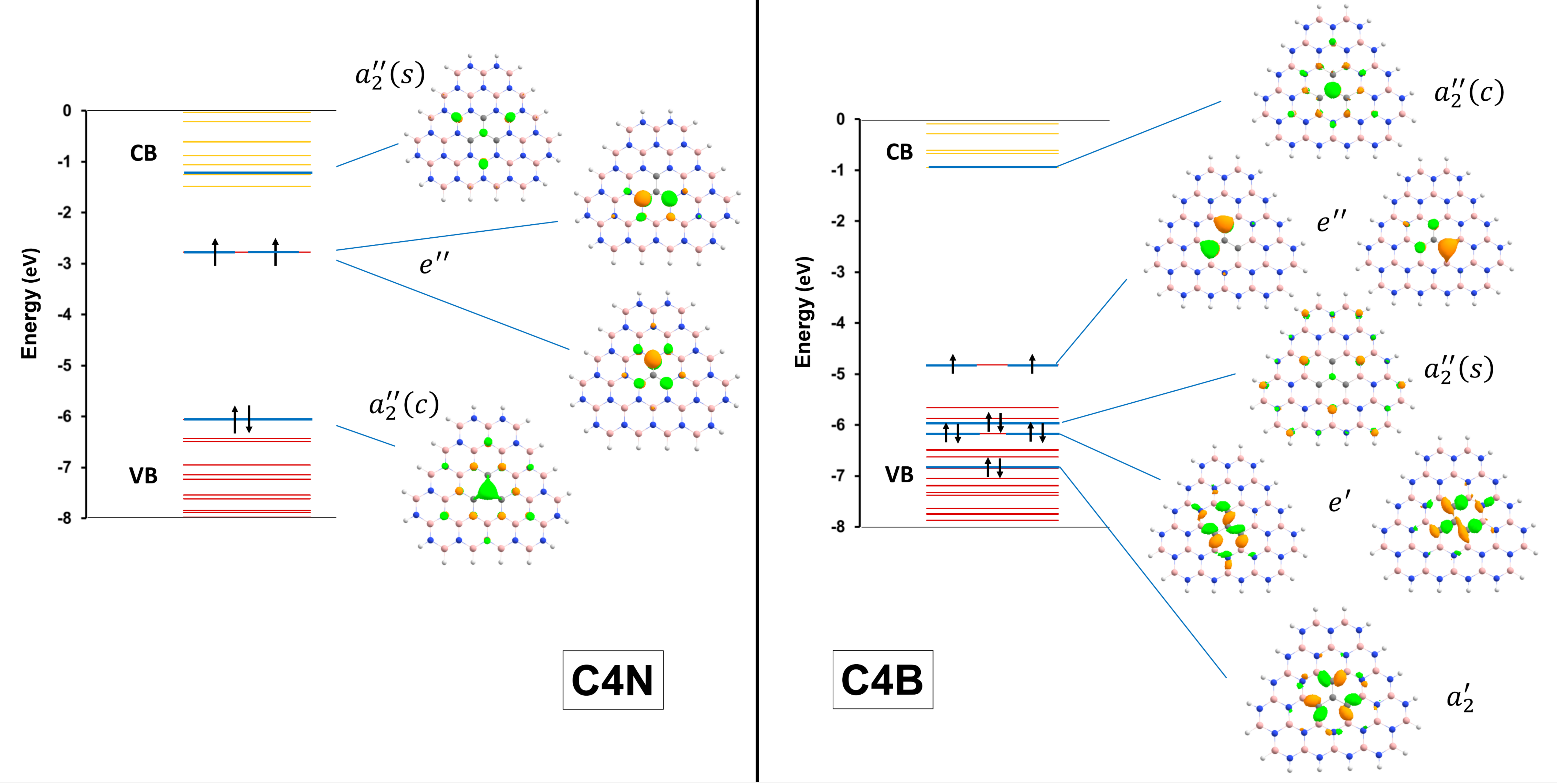}
	\caption{Orbital diagram of C4N and C4B defects, (computed at the ground state geometry of  FLAKE 2 model, ROKS-PBE/cc-pVDZ level of theory), shown together with the isosurface and symmetry of C4 centered defect molecular orbitals.  }
        \label{fig:supplementary-orbital_diagram}  
\end{center}
\end{figure}

\subsection{Identification of excited states (TD-DFT)}

To identify the lowest-lying excited electronic states, we carried out  time-dependent density functional theory (TD-DFT)\cite{Petersilka_1996}  calculations  on the ground state geometry of FLAKE 1 using the ORCA program suite. 10 triplet roots and 10 spin-flipped singlet roots were requested at TD-PBE0/cc-pVDZ level of theory. 
Even though such a calculation is not expected to provide accurate energies (and electronic properties),  it still reliably provides the most relevant orbital-to-orbital excitations to be studied.  
In Table \ref{tab:tddft_energies}, we summarize the C4 center related excitations found among the 10 roots. (The rest of the obtained roots typically describe electron transitions to/from rather delocalized orbitals belonging to the conduction/valence band.) 

\begin{table}[H]
\begin{center}
\caption{\label{tab:tddft_energies} Composition and energy of excited states based on TD-PBE0/cc-pVDZ calculations on the ground state geometry.}
 \begin{tabular}{c|c|c|ccc}
 \hline
 Defect & Leading excitation(s) (weight) & Degeneracy & Identified state & Relative energy (eV) \\ \hline
\multirow{4}{*}{C4N}   & Ground state & 1 & $^{3}A_2^\prime$ & 0.00\\
& $e^{\prime\prime}[\alpha] \rightarrow  e^{\prime\prime}[\beta]$ (88\%) & 2 & $^{1}E^\prime$ & 1.18\\
& $e^{\prime\prime}[\alpha] \rightarrow e^{\prime\prime}[\beta]$ (98\%) & 1 & $^{1}A_1^{\prime}$ & 1.54\\
& $e^{\prime\prime}[\alpha] \rightarrow a_2^{\prime\prime}(s)[\alpha]$ (74\%)& 2 & $^{3}E^{\prime}$ & 2.22\\
\hline
\multirow{6}{*}{C4B}   & Ground state & 1 & $^{3}A_2^\prime$ &  0.00\\
& $e^{\prime\prime}[\alpha] \rightarrow  e^{\prime\prime}[\beta]$ (92\%) & 2 & $^{1}E^\prime$ & 1.37\\
& $e^{\prime\prime}[\alpha] \rightarrow e^{\prime\prime}[\beta]$ (99\%) & 1 & $^{1}A_1^{\prime}$ & 1.90\\ 
& $e^{\prime}[\beta] \rightarrow e^{\prime\prime}[\beta]$ (91\%) &  1 & $^{3}A_2^{\prime\prime}$  & 3.15\\
& $e^{\prime}[\beta] \rightarrow e^{\prime\prime}[\beta]$ (73\%), $a_2^{\prime}[\beta] \rightarrow e^{\prime\prime}[\beta]$ (19\%) & 2 & $^{3}E^{\prime\prime}$ & 3.15\\
&$a_2^{\prime\prime}(s)[\beta] \rightarrow e^{\prime\prime}[\beta] (81\%)$ & 2 & $^{3}E^{\prime}$ & 3.27\\ 

\hline
 \end{tabular}
\end{center}
\end{table}

\subsection{Choice of active orbitals by orbital analysis of the DMRG solution}
\label{sect:dmrg}

To identify the chemically most relevant orbitals  of the lowest vertical excitations in a "reference-free" unbiased manner, we performed exploratory density matrix renormalization group (DMRG)\cite{White-1999} calculations which is a  multireference wave-function method capable to treat several dozens of strongly correlated orbitals\cite{Szalay-2015a,Olivares-2015,Baiardi2020}.

We studied the full valence space of FLAKE 1 of C4N and C4B defects. In this model $\sigma_{sp^2}$,$\sigma*_{sp^2}$, $\pi_{p_z}$ and $\pi*_{p_z}$ orbitals between B, C and N atoms were considered which were selected from  the full set of  spin-restricted open-shell  Kohn-Sham orbitals following Pipek-Mezey orbital localization\cite{Pipek-1989}. We note that terminating N-H and B-H bonds were omitted from the simulations as they are not relevant in the understanding of the defect embedded in macroscopic hBN layer. 

Accordingly, for both C4N and C4B, we correlated 76 electrons on  76 orbitals in the DMRG simulations considering the complete active space (CAS) protocol\cite{Roos1987}. Based on the state-averaged mutual information pattern of the target states, which had been obtained by cheap preliminary DMRG calculations, we determined an optimal DMRG orbital ordering for both defects\cite{barcza_2011}.
In the warmup sweep of the DMRG, environmental were constructed according to the restricted active space protocol in order to retrieve correlation effects of all orbitals  from the initial DMRG steps\cite{barcza_2022b}.
The DMRG truncation is based on the spectrum of the state-averaged reduced density matrix.
The relevant orbitals of the excitations, can already be predicted from  DMRG calculation with low bond dimension\cite{stein_2016,barcza_2022a}. Correspondingly,  in order to profile the convergence of the relative energies, we performed DMRG simulations with fixed $M=100,200,500$ number of block states.
Also note that  we found that the relative  DMRG spectrum predicted for $M$ number of retained states reaches chemical accuracy within 5 DMRG sweeps.

\begin{table}[H]
\begin{center}
\caption{\label{tab:CAS_DMRG_energy} Vertical energies  of FLAKE1 obtained in CAS based description. DMRG simulations were
 performed on the valence space of 76 spatial orbitals and 76 electrons for increasing $M$ block states.
All values were calculated at the geometry of the ground state ($^{3}A_2^\prime$) of FLAKE1 in ROKS basis, and are presented in eV units. }
 \begin{tabular}{c|c|ccc}
 \hline
 Defect & Electronic state & DMRG(M=100)  & DMRG(M=200)  & DMRG(M=500)  \\ \hline
\multirow{4}{*}{C4N}   & $^{3}A_2^\prime$ & 0.00 & 0.00 & 0.00  \\ 
& $^{1}E^\prime$ &  0.85 & 0.84 & 0.80\\
& $^{1}A_1^\prime$ &    2.50 & 2.38 & 2.27\\
& $^{3}E^\prime$ &  4.67 & 4.56 & 4.32\\ 
\hline
\multirow{6}{*}{C4B}   & $^{3}A_2^\prime$ & 0.00 & 0.00 & 0.00 \\ 
& $^{1}E^\prime$ &  1.35 & 1.27 & 1.23\\
& $^{1}A_1^\prime$ &    4.05 & 3.86 & 3.45 \\
& $^{3}A_2^{\prime\prime}$  & 5.11 &  5.02 & 4.96\\
& $^{3}E^{\prime\prime}$ &  4.87 & 4.82 & 4.45 \\
& $^{3}E^\prime$ &   5.02 &  4.87 & 4.58 \\ \hline
 \end{tabular}
 \end{center}
\end{table}

In line with the previous TD-DFT results, we requested 3 singlet and 3 triplet roots for C4N, while 3 singlet and 6 triplet roots for C4B. The DMRG energies are summarized in Table~\ref{tab:CAS_DMRG_energy}. 
We found that increasing the number of block states by a factor of five (M=100 vs 500), the excitation energies typically decrease only by 1-10\%.

\begin{figure}[H]
\begin{center}
	\includegraphics[width=0.49\textwidth]{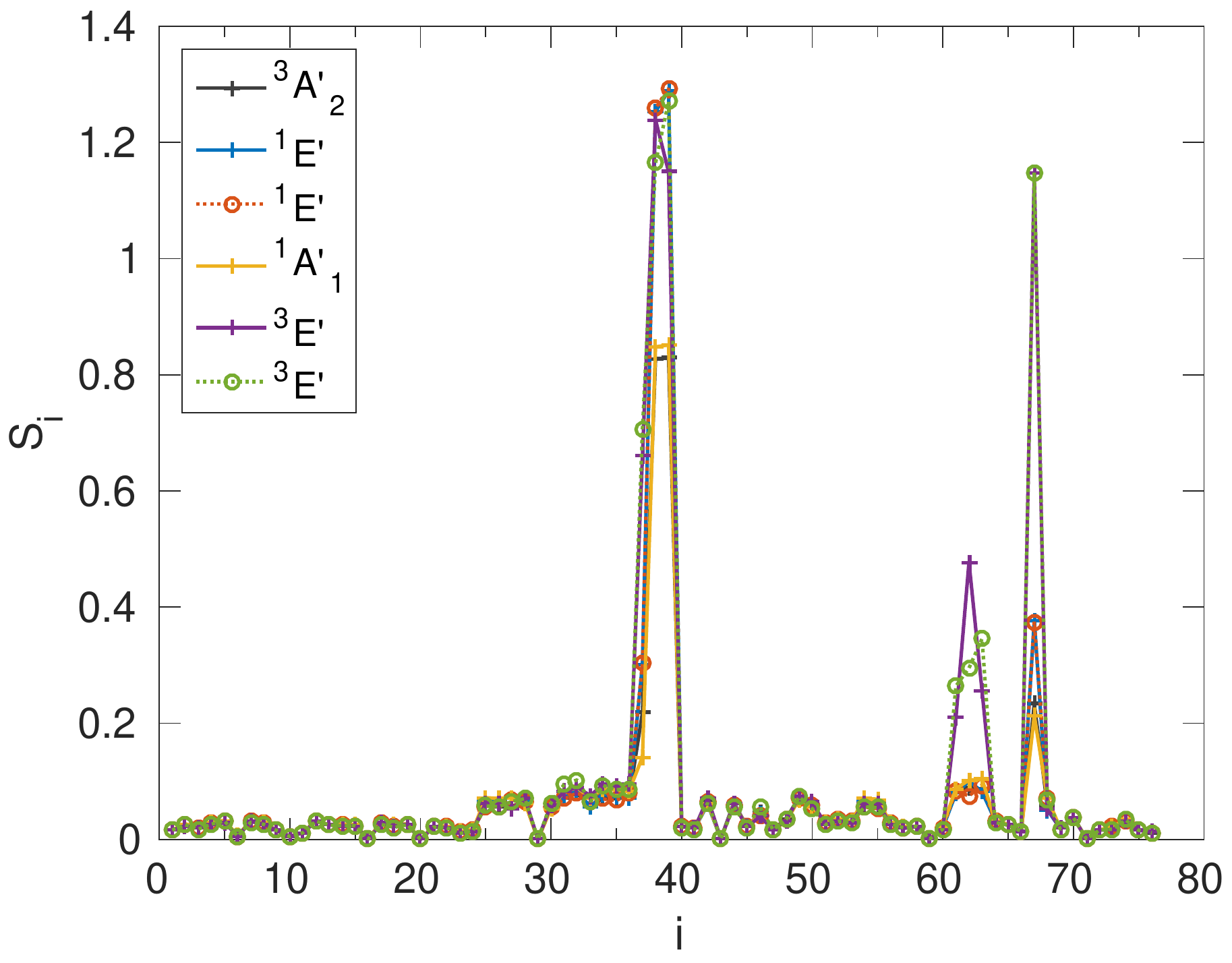}
    \includegraphics[width=0.49\textwidth]{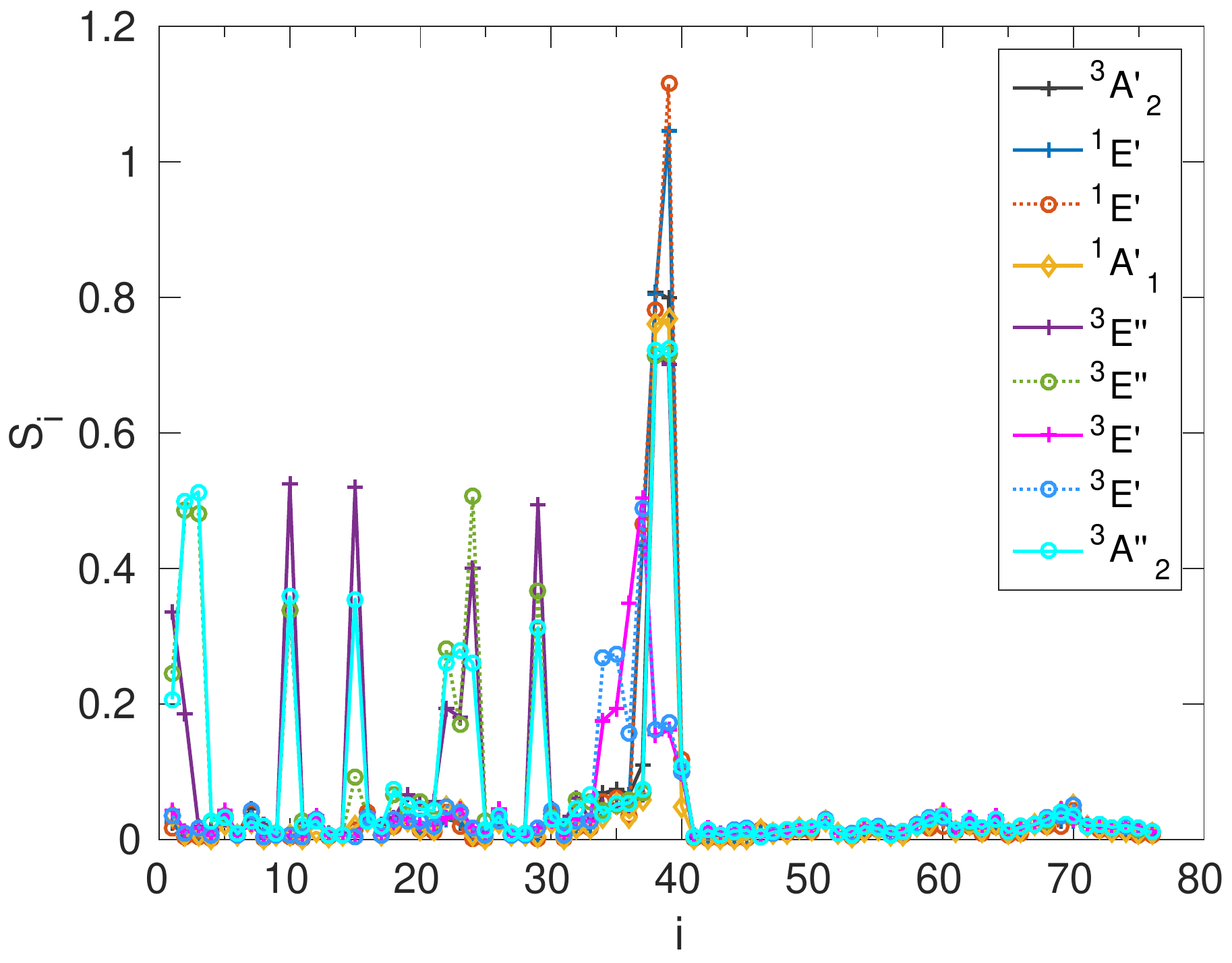}
 \caption{(Left) Single orbital entropy $S_i$ of each target state of the FLAKE1 C4N defect obtained by DMRG keeping M=100 block states.
 (Right) Similar figure but for FLAKE1 C4B. }
	\label{fig:DMRG_entropy}  
\end{center}
\end{figure}

\begin{figure}[H]
\begin{center}
\includegraphics[width=\textwidth]{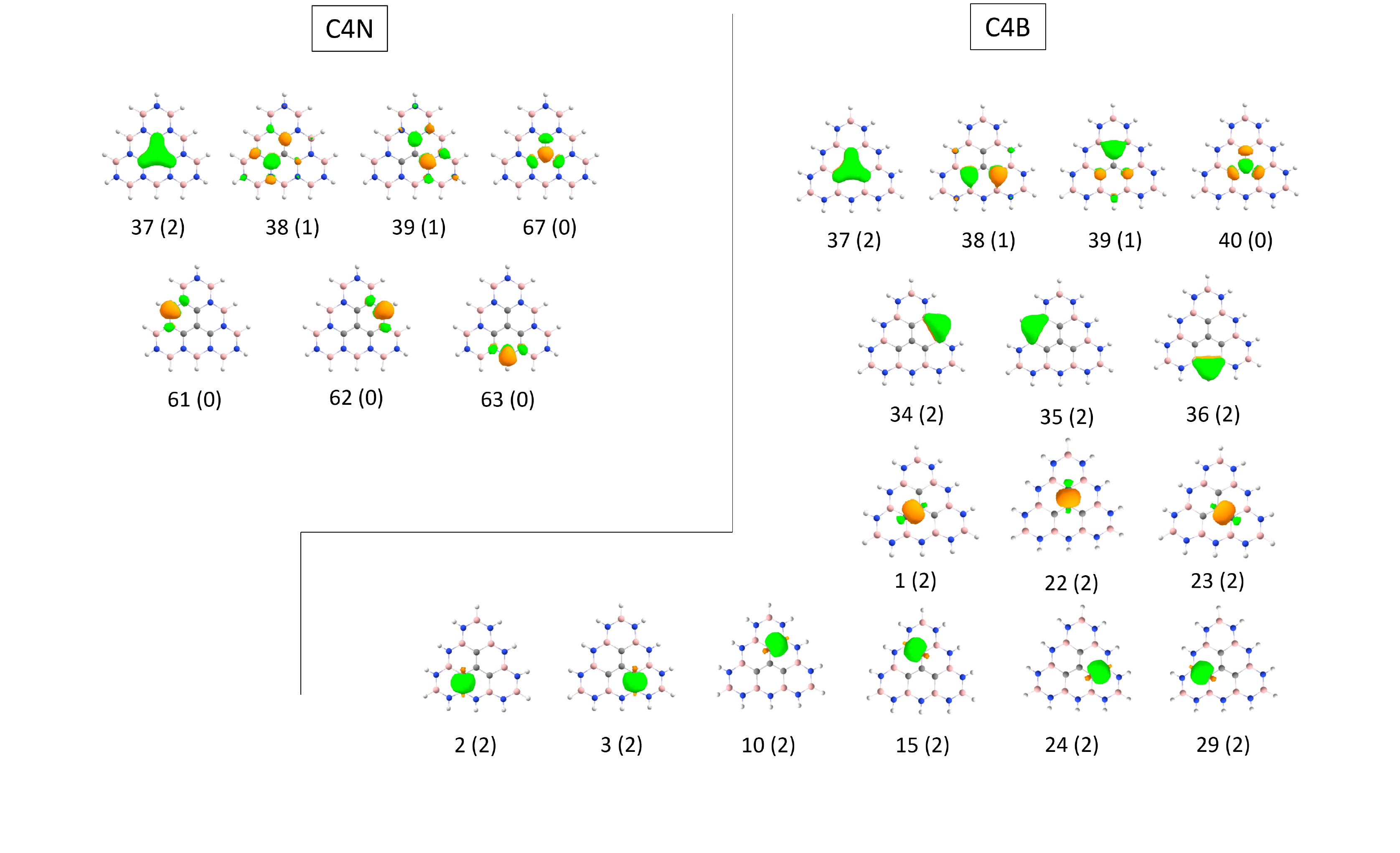}
 \caption{Shape of localized DFT orbitals of FLAKE 1 selected for CASSCF active space.  For better visibility, the presented orbitals are grouped according to their characteristic structure. Number of assigned electrons are shown in parentheses.}
	\label{fig:active_orbitals}  
\end{center}
\end{figure}

The chemical relevance of the molecular orbitals was assessed by the  single-orbital entropy profile\cite{barcza_2011,stein_2016} for each electronic state of interest. 
We found that for C4N defect the obtained DMRG roots were found to characteristically correspond to the TD-DFT electronic states of Table~\ref{tab:tddft_energies}, here we identified the  seven $p_z$ orbitals localized on the defect carbons (indexed 37, 38, 39, 67) and the neighboring  borons (61, 62, 63) with substantial orbital entropy, see left panel of Fig.\ref{fig:DMRG_entropy} and Fig.\ref{fig:active_orbitals}.
In the case of C4B, owing to the weaker bonding of the carbons to the neighboring electron-deficient borons, we found that $\sigma$ system also plays substantial role, i.e., besides the $p_z$ orbitals on the carbons (37, 38, 39, 40) and on inner N atoms (34, 35, 36) $\sigma$ orbitals forming C-C (1, 22, 23) and neighboring B-C (2, 3, 10, 15, 24, 29) bonds have outstanding entropy contribution, see right panel of Fig.\ref{fig:DMRG_entropy} and Fig.\ref{fig:active_orbitals}. 

Accordingly, we selected a 4-electron 7-orbital active space for C4N, and a 28-electron 16-orbital active space for C4B. 


\subsection{Choice of the size of the flake model (CASSCF-NEVPT2)}
The single-point electronic properties  discussed in the paper were determined at CASSCF/cc-pVTZ level of theory while the  single-point energies are corrected by the second-order N-electron valence perturbation theory (NEVPT2)\cite{nevpt2}. State-average calculations were performed, requesting multiple singlet and triplet roots.
The active space was selected from localized PBE0/cc-pVTZ molecular orbitals corresponding to density matrix renormalization group predictions as described in Sect.~\ref{sect:dmrg}. 

According to preliminary CASSCF-NEVPT2 calculations with different flake model sizes (Table \ref{tab:flake_convergence_casscf-nevpt2}), the  relative energy of the lowest-lying  vertical electronic states already reaches convergence at the size of FLAKE 2. 
On the other hand, it was previously shown that the calculation of phonons is more sensitive to the system size and requires a sufficiently large model~\cite{reimers_photoluminescence_2020}. Thus, we decided on computing the vibrational spectrum of the defect based on FLAKE 3, while other (relaxed) molecular properties were obtained using FLAKE 2.

\begin{table}[H]
\begin{center}
\caption{\label{tab:flake_convergence_casscf-nevpt2} Vertical energy spectrum obtained CASSCF-NEVPT2/cc-pVTZ energy levels for increasing flake sizes. Note that DLPNO-NEVPT2 perturbation\cite{dlpno-nevpt2} was applied for FLAKE 3 due to the unreasonably large computational cost of non-approximated NEVPT2. The active space ((4,7) or (28,16) for C4N and C4B defects, respectively) was selected from the set of localized Kohn-Sham orbitals as described in the previous section. All values were calculated at the geometry of the ground state ($^{3}A_2^\prime$), and are presented in eV units. (See Fig. \ref{fig:flakes} for the visualization of the flake models.)  As the difference between FLAKE 2 and FLAKE 3 excitation energies remains below 0.15 eV (i. e. the expected error margin of CASSCF-NEVPT2) in all cases, FLAKE 2 can be considered as a reasonable model for computing energies and electronic molecular properties. } 
 \begin{tabular}{c|c|ccc }
 \hline
 Defect & Electronic state & FLAKE 1 & FLAKE 2 & FLAKE 3  \\ \hline
\multirow{4}{*}{C4N}   & $^{3}A_2^\prime$ & 0 & 0  & 0   \\
& $^{1}E^\prime$ & 0.64 & 0.70 & 0.69   \\
& $^{1}A_1^\prime$ & 1.18 & 1.22 & 1.09  \\
& $^{3}E^\prime$ & 2.32 & 2.94 & 3.06   \\
\hline
\multirow{6}{*}{C4B}  & $^{3}A_2^\prime$ & 0 & 0 & 0 \\ 
& $^{1}E^\prime$ &  0.76 & 0.75 & 0.72   \\
& $^{1}A_1^\prime$  & 1.17 & 1.08 &  1.04 \\
& $^{3}A_2^{\prime\prime}$  & 3.17 & 2.78 &  2.65\\
& $^{3}E^{\prime\prime}$ & 3.10 & 2.78 &  2.67 \\
& $^{3}E^\prime$ & 3.66 & 3.71 &  3.66  \\ \hline
 \end{tabular}
 \end{center}
\end{table}

\subsection{Relaxation of excited states: geometry optimization, vibrational analysis (TD-DFT), single-point energies (CASSCF-NEVPT2)}

The identified singlet excited states ($^{1}E^\prime$, $^{1}A_1^\prime$) and the lowest-lying triplet excited states (C4N: $^{3}E^\prime$; C4B: $^{3}A_2^{\prime\prime}$, $^{3}E^{\prime\prime}$) were relaxed to their optimized geometry at TD-PBE0/cc-pVDZ level of theory, by following the TD-DFT root corresponding to the desired electron configuration. It is remarkable that the (TD-)DFT geometry optimization process automatically accounts for Jahn-Teller distortion effects (if applicable) - in this way, we discovered that the optimization of $^{3}A_2^{\prime\prime}$ and $^{3}E^{\prime\prime}$ gives the same geometry of C$_{2v}$ symmetry, which we handle as the fist triplet excited state of C4B in the following.   
After the optimization process, vibrational analysis (FLAKE 3, PBE0/cc-pVDZ level) and single-point energy calculations (FLAKE 2, CASSCF-NEVPT2/cc-pVTZ) were performed on the obtained relaxed structures. We note that the vibrational spectrum was calculated by freezing the N-H and B-H bonds using the partial Hessian technique\cite{partial_hess}, as these bonds are not present in actual hBN samples.  

\subsection{Photoluminescence rates and photoluminescence spectra}

The rate and spectrum of photoluminescence, corresponding to the relaxation of the triplet excited state to the triplet ground state, was calculated using the excited state dynamics (ESD) module of ORCA\cite{orca_esd}. In this framework, the dynamics are calculated from first principles, assuming harmonic nuclear movement and using the analytic solution of the path integral of the multidimensional harmonic oscillator to solve Fermi's golden rule for photon emission. In this work, we applied the adiabatic hessian model in ESD, that is, geometry and Hessian matrix for both the ground state and the excited geometry was provided from previous DFT calculations, as described above. The transition dipole moment vector and the adiabatic energy difference required for ESD was taken from the CASSCF-NEVPT2 results. The linewidth for the spectrum was set to 300 $cm^{-1}$ in order to simulate the usual experimental setup.

We note that we applied the Frank-Condon approximation, that is, the size of the transition dipole moment vector was considered to be constant and insensitive to the geometrical changes caused by vibrations.

\subsection{Spin-orbit and spin-spin couplings, zero-field splitting, intersystem crossing rates}

Spin-orbit coupling (SOC) matrix elements, spin-spin coupling (SSC) and zero-field splitting tensors were calculated in the framework of the quasi-degenerate perturbation theory (QDPT)\cite{qdpt}, in the basis of the obtained CASSCF roots. Dynamic correlation energy was taken into account by using NEVPT2 corrected diagonal energies in the QDPT matrix. 

Intersystem crossing rates can be computed analogously to photoluminescence rates as described in the previous section, except that the the probability of transition between states is described by SOC matrix elements.

\section{Comparison of periodic and flake model predictions}

\begin{table}[H]
\begin{center}
\caption{\label{tab:Bonds} Comparison of the carbon-carbon bond lengths for ground and excited state geometries calculated within different models. All values are in \AA.}
 \begin{tabular}{c|c|ccc}
 \hline
 Defect &  State &  Flake (PBE0/cc-pVDZ) &  Sheet (HSE06) & Bulk (HSE06) \\ \hline
\multirow{4}{*}{C4N} 
& $^{3}A_2^\prime$ & 1.416 & 1.415 & 1.415\\
& $^{3}E^\prime$ & 1.435, 1.436, 1.443 & 1.430, 1.430, 1.441 & 1.433, 1.434, 1.448\\
& $^{1}E^\prime$ & 1.381, 1.437, 1.437 & 1.383, 1.433, 1.434 & 1.391,1.415, 1.444\\  
& $^{1}A_1^\prime$ & 1.418, 1.418, 1.418 & - & -\\
\hline
\multirow{4}{*}{C4B}  
& $^{3}A_2^\prime$ & 1.418 & 1.408 & 1.408\\
& $^{3}A_2^{\prime\prime}$ & 1.383,1.383,1.487 & 1.368, 1.384, 1.462 & 1.405, 1.407, 1.435\\
& $^{1}E^\prime$ & 1.393,1.430,1.430 & 1.392, 1.393, 1.443 & 1.382, 1.405, 1.440\\
& $^{1}A_1^\prime$ & 1.416,1.416,1.416 & - & -\\
\hline
 \end{tabular}
\end{center}
\end{table}

\begin{table}[H]
\begin{center}

\caption{\label{tab:Energy} Comparison of the zero phonon line energies calculated within different models. All values are in eV.}  
 \begin{tabular}{c|c|ccc}
  \hline
  Defect & Transition &  Flake (CASSCF/cc-pVTZ) &  Sheet (HSE06) & Bulk (HSE06) \\ \hline
\multirow{2}{*}{C4N}  
& $^{3}A_2^\prime$ $\rightarrow$ $^{3}E^\prime$  & 2.32 & 1.98 & 1.99\\
& $^{1}E^\prime$ $\rightarrow$ $^{1}A_1^\prime$ & 0.52 & - & -\\
\hline
\multirow{2}{*}{C4B}  
& $^{3}A_2^\prime$ $\rightarrow$ $^{3}A_2^{\prime\prime}$ & 2.37 & 2.40 & 2.44\\
& $^{1}E^\prime$ $\rightarrow$ $^{1}A_1^\prime$ & 0.35 & - & -\\
 \hline
\end{tabular}
  \end{center}
\end{table}

\section{Supplementary figures and tables}

\subsection{Excited state distortion in single layer models (C4N defect)}
\label{sec:supplementary-exc_distortion} 

\begin{figure}[h!]
\begin{center}
	\includegraphics[width=0.80\columnwidth]{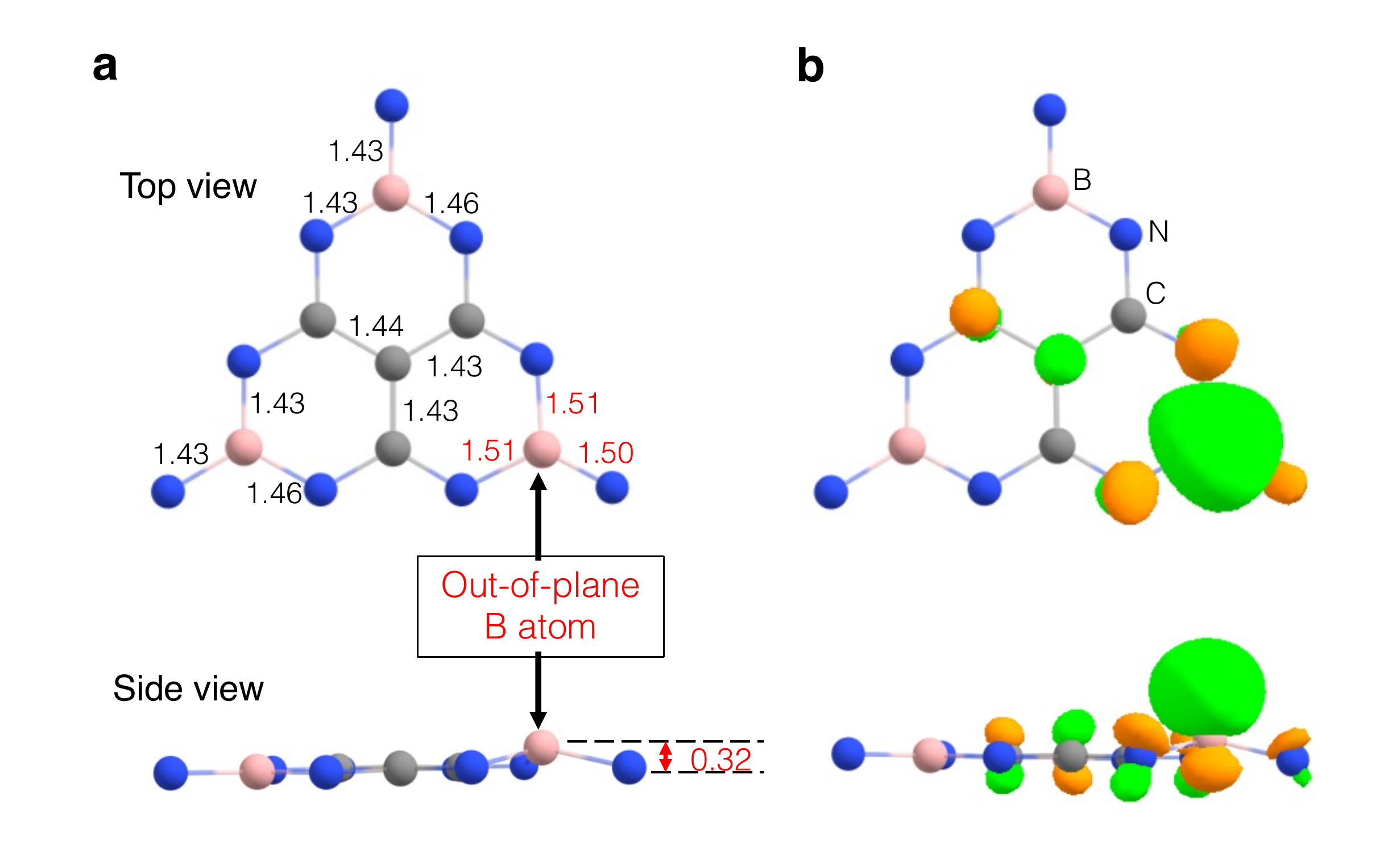}
	\caption{ Excited state configurations of the C4N defect in single layer hBN. \textbf{a} Structure of the polaronic-like distortion in the $^3E^{\prime}$ excited state in single layer hBN. \textbf{b} Localization of the highly distorted $a_2^{\prime\prime}(h)$ single particle state on the out-of-plane boron atom. }
	\label{fig:sb}  
\end{center}
\end{figure}

We note that we observe different geometry distortions for the triplet excited state of the C4N defect  in multi and single layer models. In the latter case, the highly symmetric structure spontaneously distorts beyond the Jahn-Teller effect and relaxes into a polaronic-like state where one of the second nearest neighbor boron pops out of the plane, see Fig.~\ref{fig:sb}a. In  this configuration the partially occupied $a_2^{\prime\prime}(s)$ state gets large distorted and localizes only on the p$_z$ orbital of the out of plane boron atom, Fig.~\ref{fig:sb}b. These results are consistently obtained in our periodic hybrid-DFT and CASSCF-NEVPT2 calculations. Using the latter method, we obtain an energy for the polaronic-like distortion $\sim0.5$~eV lower than the symmetric excited state configuration. On the other hand, the out-of-plane relaxation becomes energetically unfavourable in bulk and multi layer structures. Since the multi-layer  configuration is currently more relevant for the applications than the single layer configuration, we study the former in more details.

\subsection{Flakes buckled due to strain (C4N defect)}
\label{sup:sec:buckled_geom}
\begin{figure}[H]
\begin{center}
	\includegraphics[width=\textwidth]{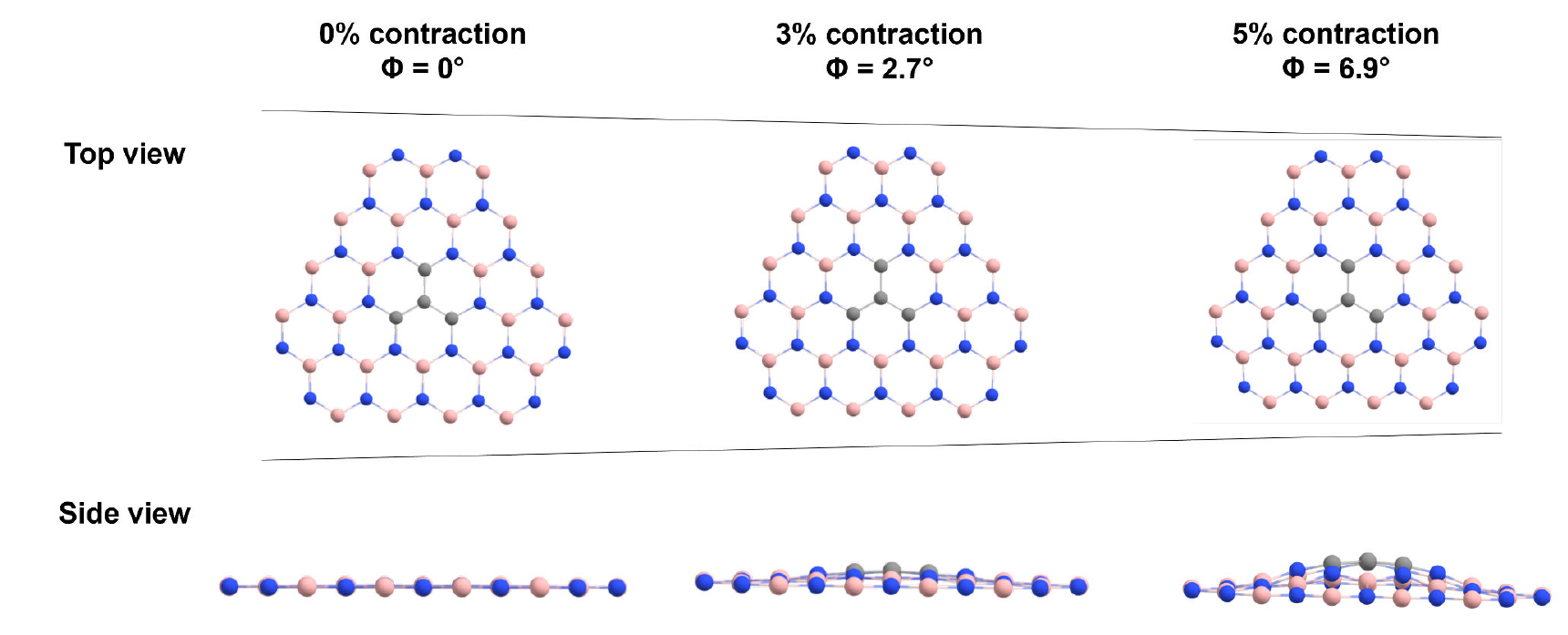}
	\caption{Visualization of the geometry of the ${}^{3}A^{'}_2$ electronic state at different stages of in-plane compression. Left: equilibrium geometry. Middle: 3\% contraction (i.e. distances between the central carbon and the outside atoms reduced by 3\% compared to equilibrium). Right: 5\% contraction (i.e. distances between the central carbon and the outside atoms reduced by 5\% compared to equilibrium). $\phi$ refers to the dihedral angle given by the position of the 4 carbon atoms. The geometries are presented using the molecular model FLAKE 2 (see \ref{fig:flakes}). Hydrogen atoms are omitted for clarity.}
        \label{fig:buckled_flakes}  
\end{center}
\end{figure}

\subsection{Vibrational modes coupled to the PL transition (C4N defect)}
\label{sup:sec:modes}
\begin{figure}[H]
\begin{center}
	\includegraphics[width=\columnwidth]{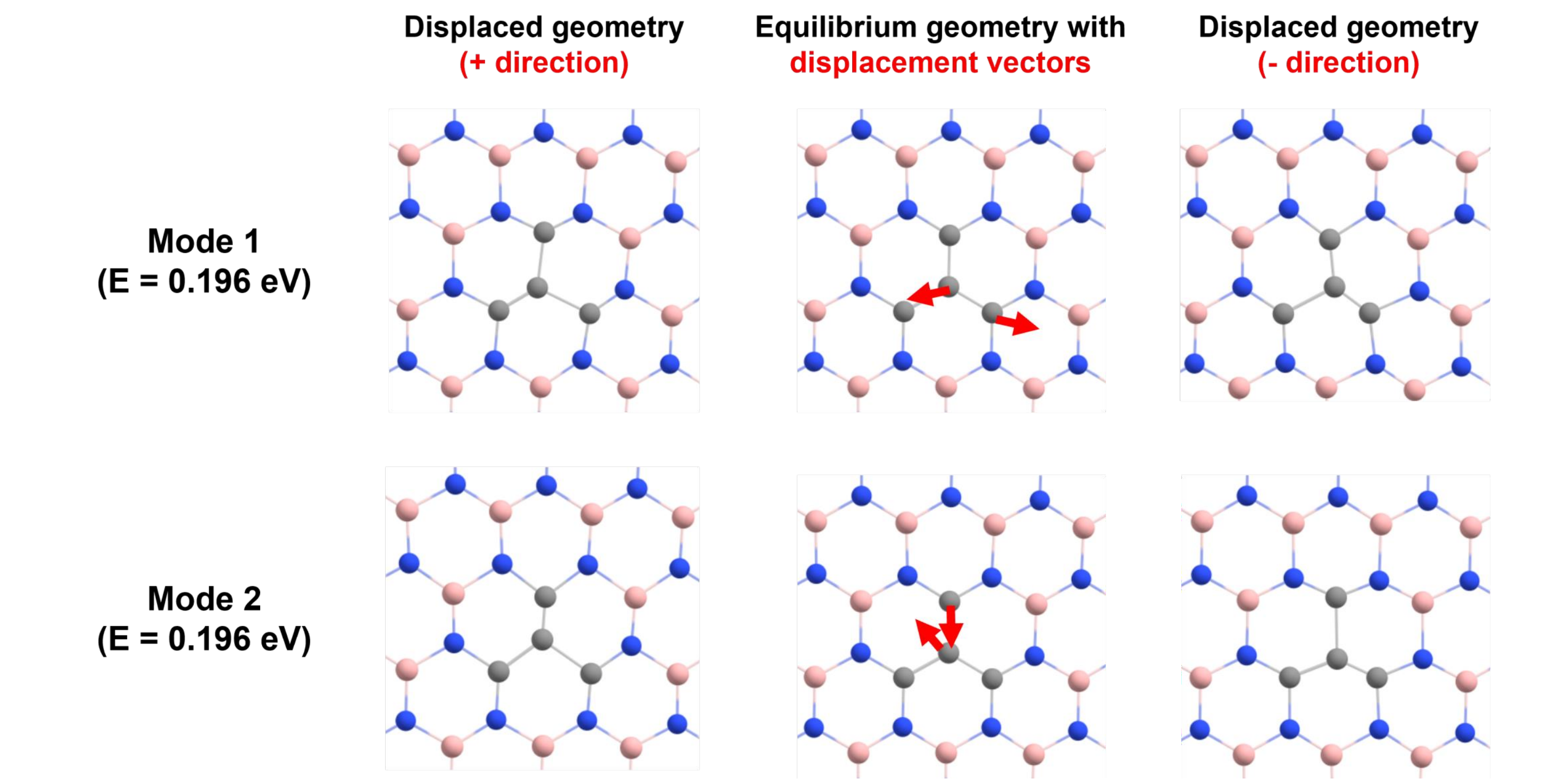}
	\caption{Supplement to Fig. 4.: Visualization of the dominant vibrational modes of the phonon sideband of ${}^{3}E^{'} \rightarrow {}^{3}A^{'}_2$ transition}
        \label{fig:modes}  
\end{center}
\end{figure}

\subsection{Estimated PL and ISC rates for buckled flakes - demonstration of spin polarization and ODMR contrast (C4N defect)}
\label{sup:sec:transitionrates}
\begin{figure}[H]
\begin{center}
	\includegraphics[width=0.8\columnwidth]{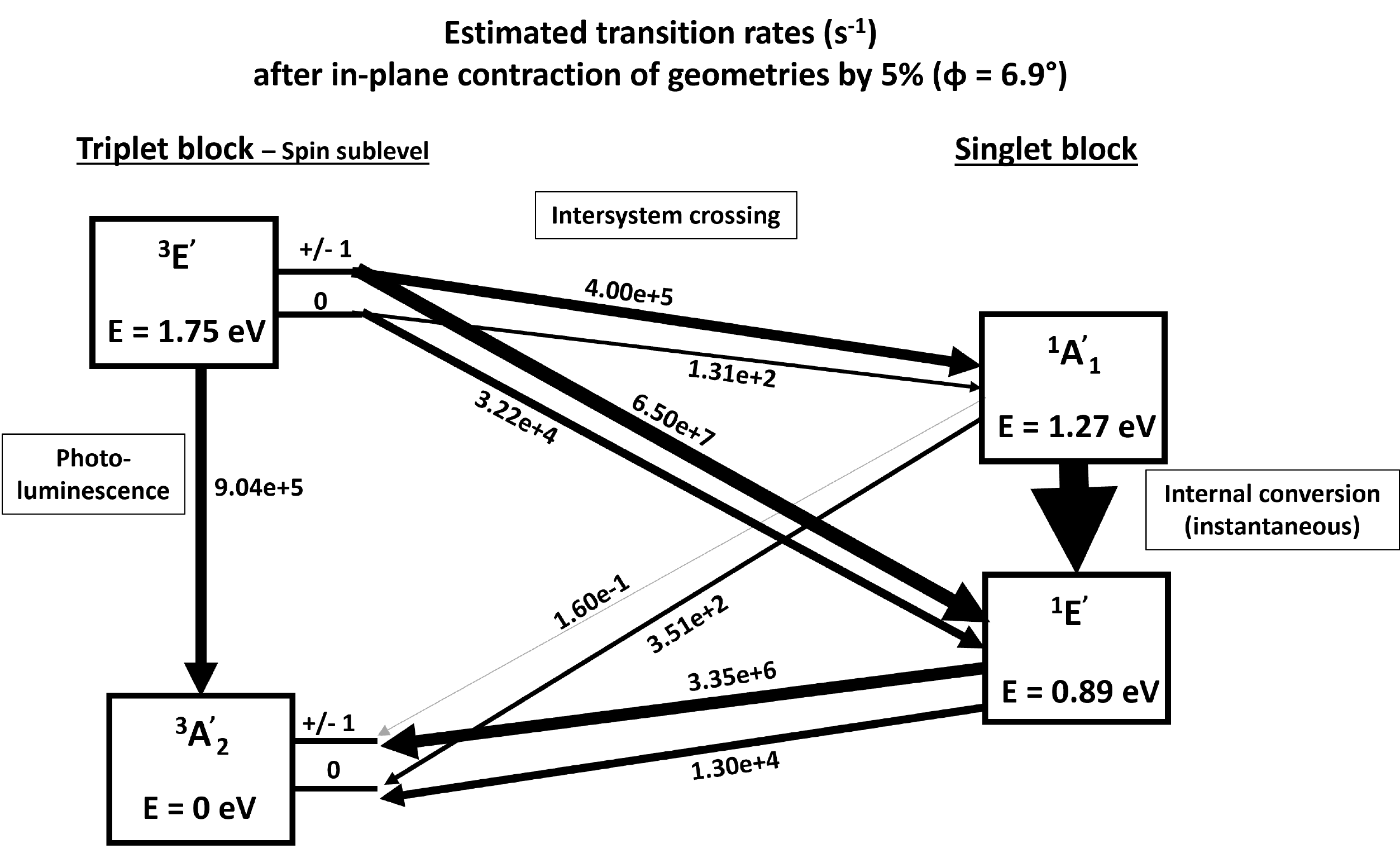}
	\caption{Estimated transition rates between the electronic states of the buckled flake ($\phi$=6.9°). The rates were computed using the Excited State Dynamics module of ORCA, based on the CASSCF(4,7)-NEVPT2 electronic energies shown in the Figure, the SOC matrix elements shown in Table 1 and the vibrational spectra of the flakes computed at TDDFT level. The thickness of reaction arrows is proportional to the order of magnitude of the represented transition rate. It can be observed that PL and ISC rates are comparable (enabling readout) and that the relaxation of ${}^{3}E^{'}$ through the singlet block results almost exclusively in the +/-1 spin sublevel of ${}^{3}A^{'}_2$ (enabling inicialization and manipulation).}
	\label{fig:tr}  
\end{center}
\end{figure}

\subsection{SOC matrix elements on displaced geometries (demonstration of Herzberg-Teller effect)}

\begin{table}[h!]
\begin{center}
\caption{\label{sup:tab:HTeffect} Spin-orbit coupling matrix elements (SOCMEs, absolute values) as obtained on CASSCF-NEVPT2 level of theory for equilibrium (${^3}A_2^{\prime}$) and displaced geometries. The displacement is given relative to the amplitude of the normal vibrational mode visualized in Figure \ref{sup:fig:out-of-plane-vibration}, displacement 0.00 denotes the equilibrium. All energy values are in GHz. Highlighted are the trends indicating the possibility of Herzberg-Teller transitions.}
 \begin{tabular}{|c|ccc|}
 \hline
 \multirow{2}{*}{  Transition} & \multicolumn{3}{c|}{Displacement }  \\ \cline{2-4}
 &  0.00 & 0.05  & 0.1 \\\hline
 \bf{$^3E^{\prime}(m_s = \pm1) \rightarrow {^1}A_1^{\prime}$} & \bf{0} & \bf{5.70} & \bf{9.81} \\
 $^3E^{\prime} (m_s = 0) \rightarrow {^1}A_1^{\prime}$ & 0  & 0 & 0   \\
 \bf{$^3E^{\prime} (m_s = \pm1) \rightarrow {^1}E^{\prime}$} & \bf{0} &  \bf{7.77} & \bf{12.99}   \\
 $^3E^{\prime} (m_s = 0) \rightarrow {^1}E^{\prime}$ & 0.27 & 0.24 & 0.27  \\
 $^1A_1^{\prime} \rightarrow {^3}A_2^{\prime} (m_s = \pm1)$ & 0 & 0 & 0  \\
 $^1A_1^{\prime}\rightarrow {^3}A_2^{\prime} (m_s = 0)$ & 2.82 & 1.98 &  1.26 \\
 \bf{$^1E^{\prime} \rightarrow {^3}A_2^{\prime} (m_s = \pm1)$} & \bf{0} & \bf{3.81} &  \bf{6.24}  \\
 $^1E^{\prime} \rightarrow {^3}A_2^{\prime} (m_s = 0)$ & 0  & 0 & 0  \\ \hline
 \end{tabular}
\end{center}
\end{table}

\begin{figure}[H]
\begin{center}
	\includegraphics[width=0.8\textwidth]{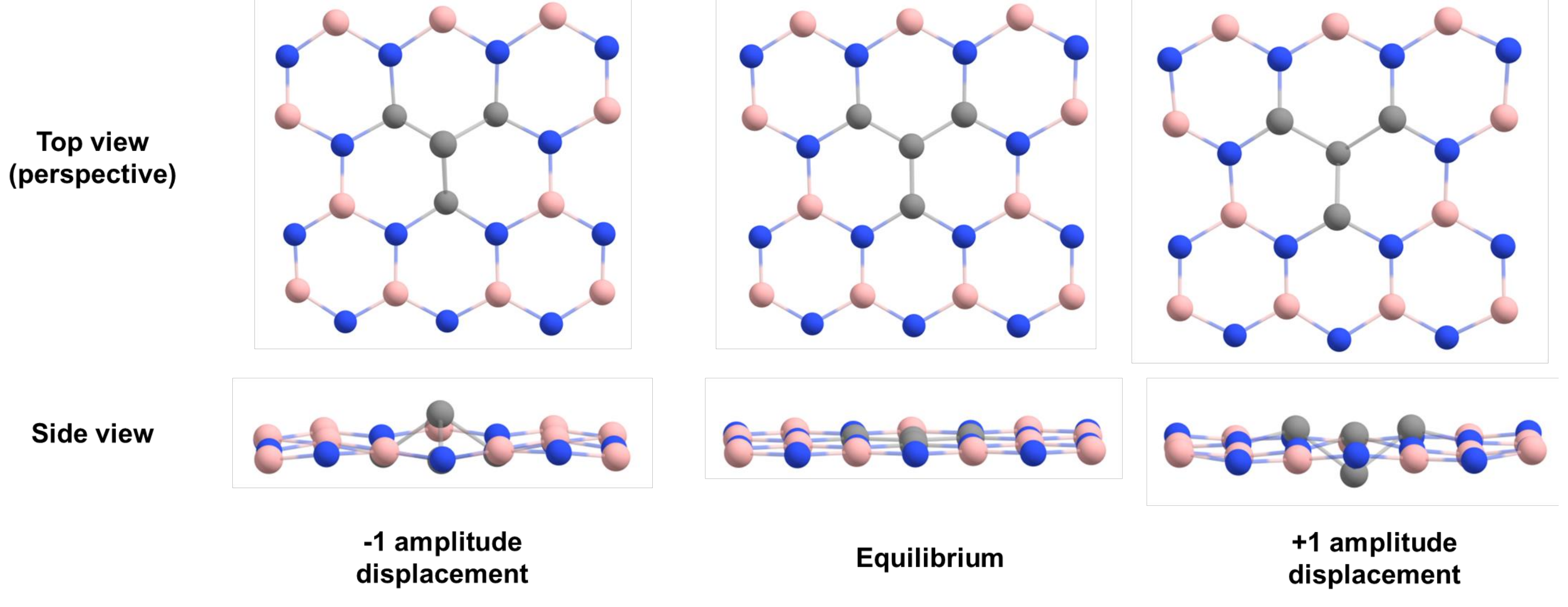}
	\caption{Visualization of the out-of plane vibrational mode of the C4 center (E = 613 $cm^{-1}$). For clarity reasons, only the center of the flake is shown.}
        \label{sup:fig:out-of-plane-vibration}  
\end{center}
\end{figure}

\subsection{Hyperfine tensors of nuclear spins}
\label{sup:sec:hyperfine}
\begin{table}[H]
\caption{\label{sup:tab:hyperfine} Supplement to Table 2:  (Left) C4N  hyperfine tensors of all coupled nuclear spins. All values are in MHz. (Right) Similar but for CBN.}
\begin{tabular}{cc}%
 \begin{tabular}[t]{|c|cccccc| }
 \hline
 site & $A_{xx}$ & $A_{yy}$ & $A_{zz} = A_z$ & $A_{xy}$ & $A_{xz}$ & $A_{yz}$ \\ 
 \hline
C0 & -33.4 & -33.4 & -57.6 & 0.0 & 0.0 & 0.0  \\ \hline
C1$_a$ & 8.7 &  8.6 & 95.0 &  -0.1  &  0.0 & 0.0 \\
C1$_a$ & 8.6 &  8.8 &  95.0 &  0.0  &  0.0 & 0.0 \\
C1$_a$ & 8.7 &  8.6 &  95.0 &   0.1  & 0.0 & 0.0 \\ 
N1$_a$ &  -3.2 &  -3.3 &  0.3  & 0.1  &  0.0  & 0.0   \\
N1$_a$ &  -3.3 &  -3.1 &  0.3  & 0.0  &  0.0  &  0.0    \\
N1$_a$ &  -3.2 &  -3.3 &  0.3  &  0.1  &  0.0  &  0.0   \\
N1$_a$ &  -3.2 &  -3.3 &  0.3  &  -0.1  &  0.0  &  0.0    \\
N1$_a$ &  -3.2 &  -3.3 &  0.3  &  -0.1  &   0.0  & 0.0    \\
N1$_a$ &  -3.3 &  -3.1 &  0.3  &  0.0  &  0.0  &  0.0    \\

B2$_a$ &  -0.3 &  -0.5  &  1.7 &  -0.1 &   0.0 &  0.0  \\
B2$_a$ &  -0.3 &  -0.5 &   1.7 &   0.1 &  0.0 &   0.0 \\
B2$_a$ &  -0.6 &  -0.3  &  1.7 &  0.0  & 0.0 &  0.0  \\

B3$_a$  &  0.2 &  -0.9  &  1.5  &  0.3   & 0.0  &  0.0 \\    
B3$_a$  &  0.2 & -0.9  &  1.5 &  -0.3  &  0.0  &  0.0 \\   
B3$_a$ &  -0.9  & 0.2  &  1.5 &  -0.3  & 0.0  &  0.0 \\    
B3$_a$ &  -0.9  &  0.2  &  1.5  &  0.3  &  0.0  &  0.0 \\   
B3$_a$ &  -0.3  & -0.3  &  1.5  &  0.7  &  0.0  &  0.0  \\
B3$_a$ &  -0.3  &  -0.3  &  1.5  & -0.7  &   0.0  &   0.0 \\ \hline
B$_A$ &  -0.3 &  -0.3  &  0.6 & 0.0 & 0.0   & 0.0 \\
B$_B$ & -0.3 &  -0.3 & 0.6 & 0.0 & 0.0 &  0.0 \\ \hline
 \end{tabular}

&

 \begin{tabular}[t]{|c|cccccc| }
 \hline
 site & $A_{xx}$ & $A_{yy}$ & $A_{zz} = A_z$ & $A_{xy}$  & $A_{xz}$ & $A_{yz}$ \\ \hline
C$_c$ & -22.9 & -22.9 & -48.0  &  0.0  &  0.0 &  0.0  \\ 
C$_{sa}$ &-2.4 &  -2.5 &  68.9  & -0.1 &  0.0  &  0.0  \\
C1$_{sb}$ &   -2.6 &  -2.6 &  68.9 &   0.0  & 0.0 &  0.0   \\
C1$_{sc}$ &   -2.4 &  -2.5 &  68.9  &  0.1  &  0.0   & 0.0 \\
B$_{1a}$ &  -10.6  & -8.5 &  -6.8  & -0.2  &  0.0  &  0.0   \\
B$_{1b}$ &   -9.2  & -9.9  & -6.8  &  1.0 &  0.0 &   0.0   \\
B$_{1c}$ &   -8.9  & -10.2  &  -6.8  &  0.8 &  0.0 &  0.0   \\
B$_{1d}$ &   -9.2  & -9.9  & -6.8 &  -1.0  &  0.0  &  0.0   \\
B$_{1e}$ &   -8.9 & -10.2 &  -6.8  & -0.8  &  0.0  &  0.0   \\
B$_{1f}$ &  -10.6  & -8.5  & -6.8  &  0.2 &  0.0  &  0.0   \\
 B$_{5a}$ &   -0.5 &  -0.9 &  -1.1  &  0.3  &  0.0  &  0.0   \\
B$_{5b}$ &   -0.5  & -0.9 &  -1.1  & -0.3  &  0.0 &  0.0   \\
B$_{5c}$ &   -1.1  &  -0.4  & -1.1  &  0.0  &  0.0  & 0.0   \\
N$_{2a}$ &    -0.3  & -0.4  &  1.5   & 0.1 &  0.0  &  0.0    \\
N$_{2b}$ &    -0.3  &  -0.4  &  1.5   & -0.1  &  0.0  &  0.0    \\
N$_{2c}$ &    -0.5  &  -0.3  &  1.5  & 0.0  &  0.0  &  0.0    \\
N$_{3a}$ &   -0.3  & -0.2  &  1.1  &   0.1  &  0.0  &  0.0    \\
N$_{3b}$ &   -0.1  &  -0.4  &  1.1  &  0.0  &  0.0 &  0.0    \\
N$_{3c}$ &   -0.1  & -0.4  &  1.1  &  0.0 &  0.0 &  0.0    \\
N$_{3d}$ &   -0.3  & -0.3  &  1.1  & -0.1  &  0.0  &  0.0    \\
N$_{3e}$ &   -0.3 &  -0.2 &   1.1   & -0.1  &  0.0 & 0.0    \\
N$_{3f}$ &   -0.3  & -0.2  &  1.1   & 0.1 &  0.0  & 0.0    \\
 B$_{6a}$ &   -0.5  & -0.4 &  -0.7  &  0.3  &  0.0  &  0.0    \\
 B$_{6b}$ &   -0.2 &  -0.7 &  -0.7  &  0.1   & 0.0  &  0.0    \\
B$_{6c}$ &   -0.2 &  -0.7 &  -0.7 &  -0.1  & 0.0 &   0.0    \\
B$_{6d}$ &   -0.5 &  -0.4 &  -0.7  & -0.3   & 0.0  &  0.0    \\
 B$_{6e}$ &   -0.7 &  -0.2  & -0.7  & -0.2  &  0.0 &  0.0    \\
B$_{6f}$ &   -0.7  & -0.2 &  -0.7  &  0.2   & 0.0  &  0.0    \\
 B$_{4a}$ &   -0.4  &  -0.8  &  -0.2  &  -0.3  &  0.0  & 0.0    \\
B$_{4b}$ &   -0.4 &  -0.8 &  -0.2 &   0.3  & 0.0  & 0.0    \\
 B$_{4c}$ &   -1.0 &  -0.2 &  -0.2 &   0.0  &  0.0  &  0.0    \\
  B$_{Ba}$ &   -0.2 &  -0.2 &   0.6  &  0.0  &  0.2  &  0.1   \\
 B$_{Bb}$ &   -0.2 &  -0.2  &  0.6 &  0.0  & -0.2  &  0.1   \\
 B$_{Bc}$ &   -0.2 &  -0.2  &  0.6  &  0.0   & 0.0 &  -0.2    \\
 B$_{Aa}$ &   -0.2 &  -0.2 &   0.6 &   0.0  & -0.2 &  -0.1    \\
 B$_{Ab}$ &   -0.2 &  -0.2  &  0.6 &  0.0   & 0.2 &  -0.1   \\
 B$_{Ac}$ &   -0.2 &  -0.2  &  0.6  &  0.0   & 0.0  &  0.2    \\
 \hline
 \end{tabular}
  \tabularnewline
\end{tabular}
\end{table}

\section{Inputs and optimized geometries}
\subsection{Sample input files}

\subsubsection{Geometry optimization and partial vibrational analysis using ORCA 5.0.3.}
\label{input_opt}

! RIJCOSX PBE0 d3bj cc-pVDZ def2/J Opt numfreq \\
\%tddft \quad \quad \quad \quad \quad \quad \quad \quad \quad \quad \#Note: the \%tddft block was omitted in the case of ground electronic states \\
nroots 10 \\
iroot [number of root of interest] \\
sf ["true" (singlet states) or "false" (triplet states)] \\
end \\
\%freq \\
partial\_hess \{ [Atom number of H atoms]\} end \\
dx 0.01 \quad \quad \quad \quad \quad \quad \quad \quad \quad \quad \#Note: Low step size in order to avoid root flipping\\
end \\
\\
* xyzfile 0 3 [Geometry guess]

\subsubsection{Single-point energy, SOC and SSC calculations using ORCA 5.0.3.}
\label{input_sp}

! cc-pvtz normalprint moread cc-pvtz/c rijcosx def2/j nevpt2\\
\%moinp [gbw file containing localized PBE0/cc-pVTZ orbitals, with active orbitals rotated together] \\
\%casscf \\
nel [number of electrons: 4 (C4N) or 28 (C4B)] \\
norb [number of active orbitals: 7 (C4N) or 16 (C4B)] \\
mult 3,1 \\
nroots [triplet and singlet roots requested: 3,3 (C4N) or 6,6 (C4B)] \\
orbstep superci \\
switchstep diis \\
shiftup 2.0 \\
shiftdn 2.0 \\
minshift 0.6 \\
maxiter 150 \\
rel \\
dosoc true \\
dossc true \\ 
printlevel 4 \\
end \\
\\
* xyzfile 0 3 [Optimized geometry]

\subsubsection{Photoluminescence rate and spectrum calculation using ORCA 5.0.3.}
\label{input_pl}

!ESD(FLUOR) NOITER \\
\%ESD \\
GSHESSIAN ["hess" file of ground state (FLAKE 3)] \\
ESHESSIAN ["hess" file of excited state (FLAKE 3)] \\
DELE [Energy difference at CASSCF-NEVPT2 level (FLAKE 2)] \\
TDIP [Transition dipole moment between the states of interest at CASSCF-NEVPT2 level (FLAKE 2)] \\
tcutfreq 100 \\
temp 0 \\
linew 300 \\
printlevel 4 \\
END \\
\\
* xyzfile 0 3 [Optimized geometry of ground state (FLAKE 3)]

\subsubsection{Intersystem crossing rate calculation using ORCA 5.0.3.}
\label{input_isc}

!ESD(ISC) NOITER \\
\%ESD \\
ISCFSHESSIAN ["hess" file of final state] \\
ISCISHESSIAN ["hess" file of initial state] \\
DELE [Energy difference at CASSCF-NEVPT2 level] \\
SOCME [SOC matrix element between the states of interest at CASSCF-NEVPT2 level] \\
coordsys cartesian \\
ifreqflag remove \\
tcutfreq 100 \\
temp 0 \\
printlevel 4 \\
END
\\
* xyzfile 0 3 [Optimized geometry of final state]

\subsection{Optimized geometries}

The geometries of molecular models (all relevant electronic states optimized at (TD-)PBE0/cc-pVDZ level) and the atomic coordinates of supercell units can be found in the supplementary file "geometries.pdf". All xyz coordinates are given in angstroems.